\documentclass[12pt]{article}
\usepackage{amssymb}
\usepackage{graphics}
\usepackage{graphicx}
\renewcommand{\theequation}{\arabic{section}.\arabic{equation}}

\begin{document}



\def\a{\alpha}
\def\b{\beta}
\def\d{\delta}
\def\e{\epsilon}
\def\g{\gamma}
\def\h{\mathfrak{h}}
\def\k{\kappa}
\def\l{\lambda}
\def\o{\omega}
\def\p{\wp}
\def\r{\rho}
\def\t{\tau}
\def\s{\sigma}
\def\z{\zeta}
\def\x{\xi}
 \def\A{{\cal{A}}}
 \def\B{{\cal{B}}}
 \def\C{{\cal{C}}}
 \def\D{{\cal{D}}}
\def\G{\Gamma}
\def\K{{\cal{K}}}
\def\O{\Omega}
\def\R{\bar{R}}
\def\T{{\cal{T}}}
\def\L{\Lambda}
\def\f{E_{\tau,\eta}(sl_2)}
\def\E{E_{\tau,\eta}(sl_n)}
\def\Zb{\mathbb{Z}}
\def\Cb{\mathbb{C}}

\def\R{\overline{R}}

\def\beq{\begin{equation}}
\def\eeq{\end{equation}}
\def\bea{\begin{eqnarray}}
\def\eea{\end{eqnarray}}
\def\ba{\begin{array}}
\def\ea{\end{array}}
\def\no{\nonumber}
\def\le{\langle}
\def\re{\rangle}
\def\lt{\left}
\def\rt{\right}

\newtheorem{Theorem}{Theorem}
\newtheorem{Definition}{Definition}
\newtheorem{Proposition}{Proposition}
\newtheorem{Lemma}{Lemma}
\newtheorem{Corollary}{Corollary}
\newcommand{\proof}[1]{{\bf Proof. }
        #1\begin{flushright}$\Box$\end{flushright}}

\baselineskip=20pt

\newfont{\elevenmib}{cmmib10 scaled\magstep1}
\newcommand{\preprint}{
   \begin{flushleft}
   \end{flushleft}\vspace{-1.3cm}
   \begin{flushright}\normalsize
   \end{flushright}}
\newcommand{\Title}[1]{{\baselineskip=26pt
   \begin{center} \Large \bf #1 \\ \ \\ \end{center}}}
\newcommand{\Author}{\begin{center}
   \large \bf
Wen-Li Yang${}^{a}$, ~Xi Chen${}^{a}$, ~Jun Feng${}^{a}$,~Kun Hao${}^{a}$,~Bo-Yu Hou${}^{a}$,\\~Kang-Jie Shi ${}^a$
 ~and~Yao-Zhong Zhang ${}^b$
 \end{center}}
\newcommand{\Address}{\begin{center}

     ${}^a$ Institute of Modern Physics, Northwest University,
     Xian 710069, P.R. China\\
     ${}^b$ The University of Queensland, School of Mathematics and Physics,  Brisbane, QLD 4072,
     Australia\\
   \end{center}}
\newcommand{\Accepted}[1]{\begin{center}
   {\large \sf #1}\\ \vspace{1mm}{\small \sf Accepted for Publication}
   \end{center}}

\preprint
\thispagestyle{empty}
\bigskip\bigskip\bigskip

\Title{
Determinant formula for the partition function of the six-vertex model with
a non-diagonal
reflecting end } \Author

\Address
\vspace{1cm}

\begin{abstract}
With the help of the F-basis provided by the Drinfeld twist or factorizing
F-matrix for the open XXZ spin chain with non-diagonal boundary
terms, we obtain the determinant representation of the partition function
of the six-vertex model  with a
non-diagonal reflecting end under domain wall boundary condition.

\vspace{1truecm} \noindent {\it PACS:} 03.65.Fd; 04.20.Jb;
05.30.-d; 75.10.Jm

\end{abstract}
\newpage
\section{Introduction}
\label{intro} \setcounter{equation}{0}
The domain wall (DW) boundary condition of a statistical model on a finite two-dimensional
lattice was introduced in \cite{Kor82} for the six-vertex model. The partition function of the model (or
DW partition function) was then given in terms of some determinant \cite{Ize87,Ize92}. Such a
determinant representation of the partition function has played an
important role in constructing norms of Bethe states, correction
functions \cite{Ess95,Kor93,Kit99} and  thermodynamical properties
of the six-vertex model \cite{Ble05}, and also in the  Toda
theories \cite{Sog93}. Moreover, it has been proven to be very
useful in solving some pure mathematical problems, such as the
problem of alternating sign matrices \cite{Kup96}. Recently, the determinant representations of the DW partition
function have been obtained for various models \cite{Car06,Fod08,Zha07}.

Since the DW partition function is calculated \cite{Ize92} as an inner product of pseudo-vacuum and some Bethe
state which is generated by pseudo-particle creation operators (given by one-row monodromy matrix related to
closed spin chain \cite{Kor93})
on the pseudo-vacuum state, the computation of the function is simplified dramatically and can be directly
calculated in the so-called F-basis \cite{Mai00} provided by the Drinfeld twist \cite{Dri83} or factorizing F-matrix.
Such a magic F-matrix has been studied extensively for other models \cite{Ter99,Alb00,Alb00-1,Alb01,Zha06,Yan06-1}.

For a two-dimensional statistical model with a reflection end \cite{Tsu98}, in addition to the local interaction vertex,
a reflecting matrix or K-matrix which describes the boundary interactions needs to be introduced
at the reflection end of the lattice (see figure 4 below). This problem is closely related to that
of open spin chain \cite{Skl88}. The DW partition function of the six-vertex model with a diagonal reflection
end was exactly calculated and expressed in terms some determinant \cite{Tsu98}.
Then such an explicit determinant representation was re-derived
\cite{Wan02, Kit07} by using F-basis of the closed XXZ chain. However, it is well known that to obtain exact solution
of open spin chain with non-diagonal boundary terms is very {\it non-trivial} \cite{Nep04,Nep03,Cao03,Yan04-1,Gie05,Yan04,Gal05,
Gie05-1,Baj06,Yan05,Doi06,Mur06,Bas07,Yan06,Gal08,Yan07,Ami10} comparing with that of the one with simple
diagonal boundary terms. In this paper, we will investigate the determinant representation of  the DW partition
function of the six-vertex model with a non-diagonal reflection end which is specified by a generic non-diagonal
K-matrix found in \cite{Veg93,Gho94}.  The result will be essential to construct the determinant representations of
scalar products of Bethe states of the open XXZ chain with non-diagonal boundary terms \cite{Yan10}.

The paper is organized as follows.  In section 2, after introducing our notation and some basic ingredients,
we construct the four boundary states which specify the DW boundary condition of the six-vertex model
with a non-diagonal reflection end. In section 3, using the vertex-face correspondence relation we express the
DW boundary partition function in terms of the particular matrix element of the product of face type
pseudo-particle creation operators. In section 4, we present the F-matrix of the
open XXZ chain with non-diagonal boundary terms and give  the completely
symmetric and polarization free representations of the
pseudo-particle creation operators  in the F-basis. In section 5,
with the help of the F-basis provided by the F-matrix we obtain the determinant representation of
the DW partition function.  In section 6, we summarize our results and give some discussions.
Some detailed technical proof is given in Appendix A.


\section{ Six-vertex model with a reflecting end}
\label{XXZ} \setcounter{equation}{0}
In this section, we briefly review the DW boundary condition for
the six-vertex model  with non-diagonal reflecting end on an $N\times 2N$ rectangular lattice.

\subsection{The six-vertex R-matrix and associated K-matrix}

Throughout, $V$ denotes a two-dimensional linear space. The
well-known six-vertex model R-matrix $\R(u)\in {\rm End}(V\otimes
V)$ \cite{Kor93} is given by \bea
\R(u)=\lt(\begin{array}{llll}a(u)&&&\\&b(u)&c(u)&\\
&c(u)&b(u)&\\&&&a(u)\end{array}\rt).\label{r-matrix}\eea The
coefficient functions read: $a(u)=1$, $b(u)=\frac{\sin u}{\sin(u+\eta)}$,
$c(u)=\frac{\sin\eta}{\sin(u+\eta)}$. Here we assume  $\eta$ is a
generic complex number. The R-matrix satisfies the quantum
Yang-Baxter equation (QYBE), \bea
R_{1,2}(u_1-u_2)R_{1,3}(u_1-u_3)R_{2,3}(u_2-u_3)
=R_{2,3}(u_2-u_3)R_{1,3}(u_1-u_3)R_{1,2}(u_1-u_2),\label{QYB}\eea
and the unitarity, crossing-unitarity and quasi-classical
properties \cite{Yan04-1}. We adopt the standard notations: for
any matrix $A\in {\rm End}(V)$ , $A_j$ (or $A^j$) is an embedding
operator in the tensor space $V\otimes V\otimes\cdots$, which acts
as $A$ on the $j$-th space and as identity on the other factor
spaces; $R_{i,j}(u)$ is an embedding operator of R-matrix in the
tensor space, which acts as identity on the factor spaces except
for the $i$-th and $j$-th ones.

For  a model with reflection end or open spin chain \cite{Skl88}, in addition to the R-matrix, one
needs to introduce  K-matrix $K(u)$ which
satisfies the reflection equation (RE)
 \bea &&\R_{1,2}(u_1-u_2)K_1(u_1)\R_{2,1}(u_1+u_2)K_2(u_2)\no\\
 &&~~~~~~=
K_2(u_2)\R_{1,2}(u_1+u_2)K_1(u_1)\R_{2,1}(u_1-u_2).\label{RE-V}\eea

In this paper, we consider the K-matrix $K(u)$ which is a
generic solution to the RE (\ref{RE-V}) associated the six-vertex
model R-matrix  \cite{Veg93,Gho94}
\bea K(u)=\lt(\begin{array}{ll}k_1^1(u)&k^1_2(u)\\
k^2_1(u)&k^2_2(u)\end{array}\rt).\label{K-matrix}\eea
The coefficient functions are \bea && k^1_1(u)=
\frac{2\cos(\l_1-\l_2) -\cos(\l_1+\l_2+2\zeta)e^{-2iu}}
{4\sin(\l_1+\zeta+u)
\sin(\l_2+\zeta+u)},\no\\
&&k^1_2(u)=\frac{-i\sin(2u)e^{-i(\l_1+\l_2)} e^{-iu}}
{2\sin(\l_1+\zeta+u) \sin(\l_2+\zeta+u)},\no\\
&&k^2_1(u)=\frac{i\sin(2u)e^{i(\l_1+\l_2)} e^{-iu}}
{2\sin(\l_1+\zeta+u) \sin(\l_2+\zeta+u)}, \no\\
&& k^2_2(u)=\frac{2\cos(\l_1-\l_2)e^{-2iu}- \cos(\l_1+\l_2+2\zeta)}
{4\sin(\l_1+\zeta+u)\sin(\l_2+\zeta+u)}.\label{K-matrix-2-1} \eea
It is
very convenient to introduce a vector $\l\in V$ associated with
the boundary parameters $\{\l_i\}$, \bea
 \l=\sum_{k=1}^2\l_k\e_k, \label{boundary-vector}
\eea  where $\{\e_i,\,i=1,2\}$ form the orthonormal basis of $V$
such that $\langle \e_i,\e_j\rangle=\d_{ij}$.


\subsection{The model}
The partition function of a statistical model on a two-dimensional
lattice  is defined by the following:
\begin{eqnarray}
 Z=\sum \exp\{-\frac{E}{kT}\}, \nonumber
\end{eqnarray}
where $E$ is the energy of the system, $k$ is the Boltzmann
constant, $T$ is the temperature of the system,  and the summation
is taken over all possible configurations under the particular
boundary condition such as the DW boundary condition. The model we
consider here has six allowed local bulk vertex configurations

\begin{center}
\includegraphics[width=0.8\textwidth]{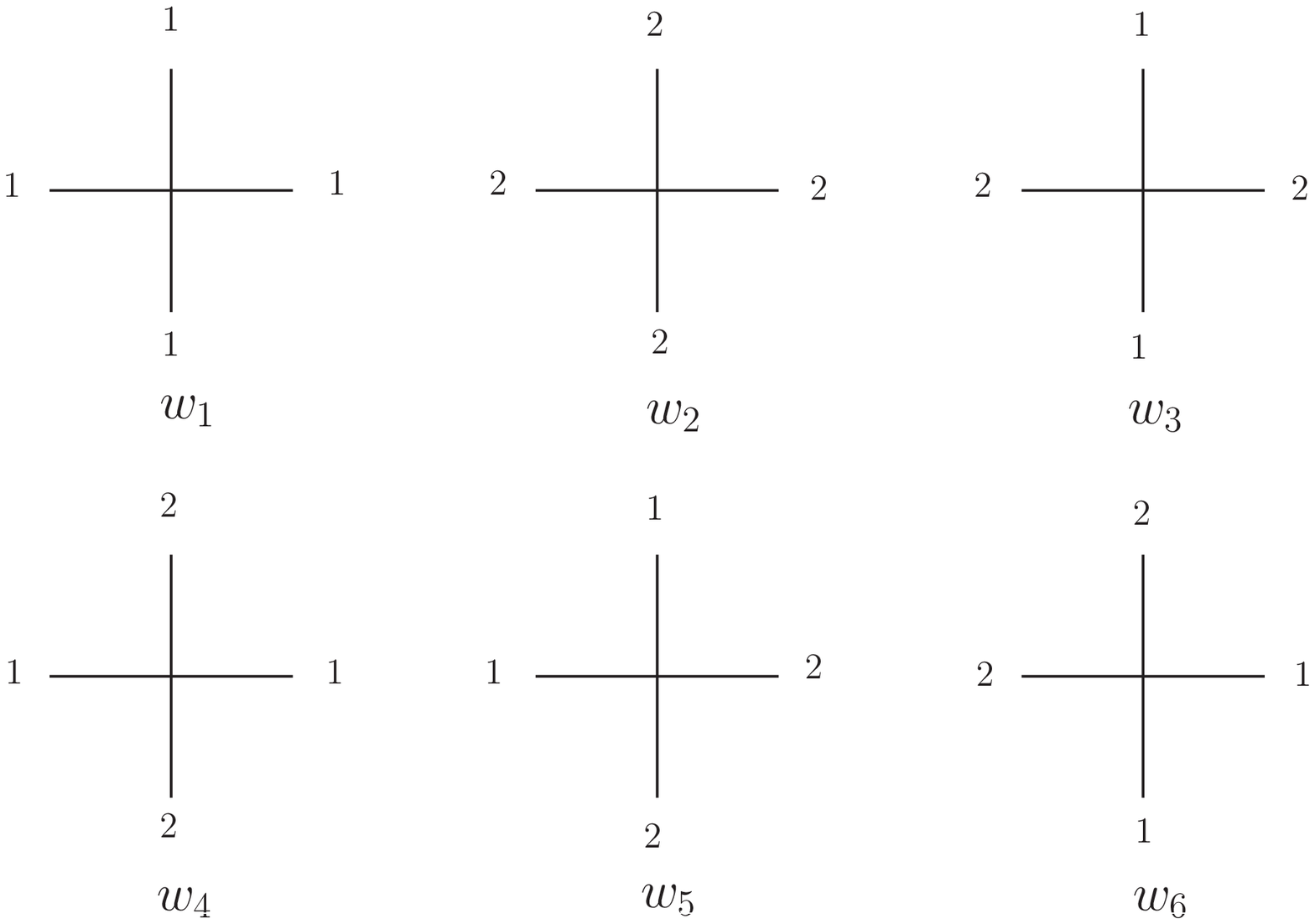}\\
{\small{\bf Figure 1.} Vertex configurations and
their associated Boltzmann weights.}
\end{center}


\noindent and four allowed configurations at each
reflection end


\begin{center}
\includegraphics[width=0.7\textwidth]{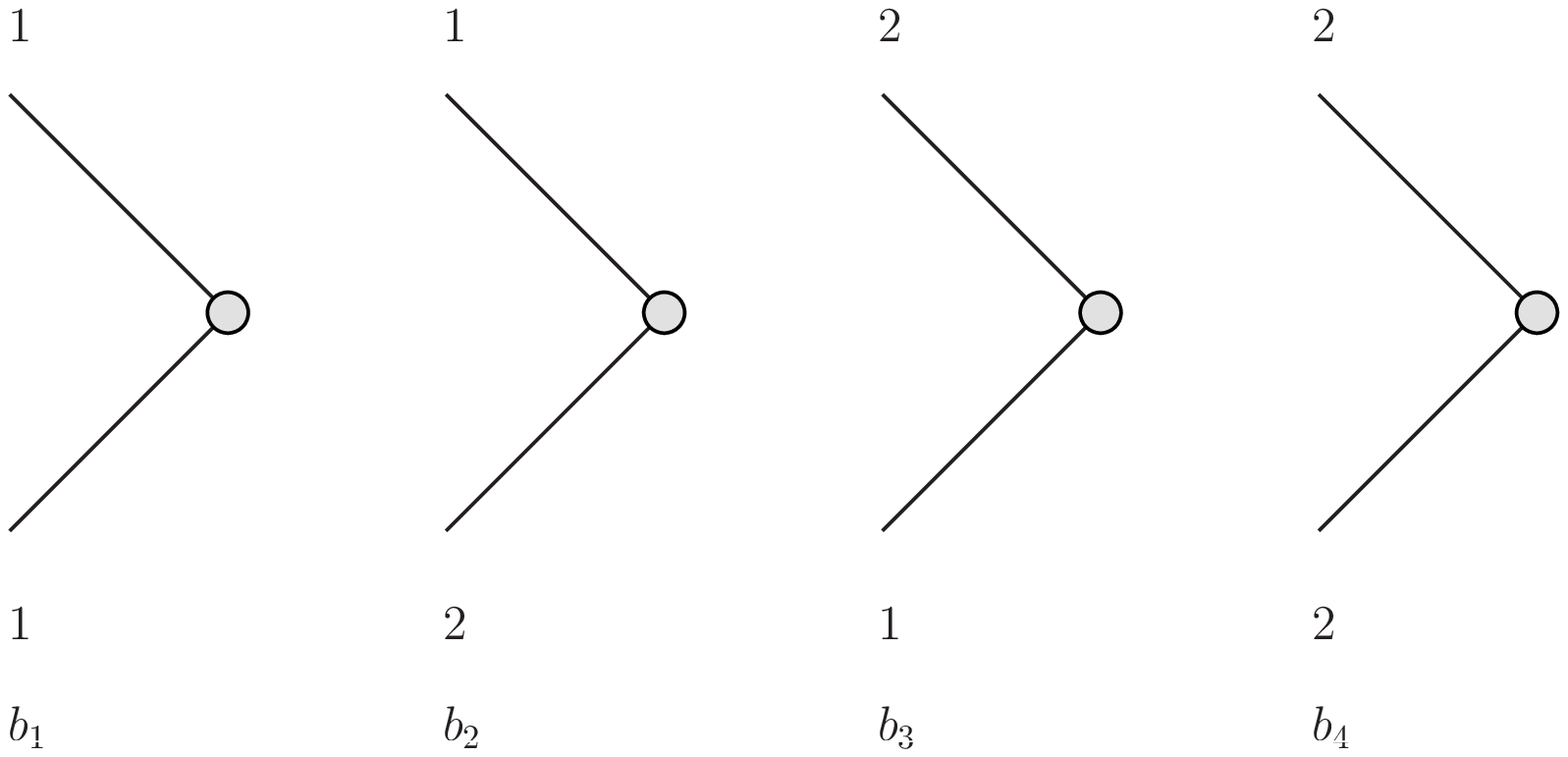}\\
{\small{\bf Figure 2.} Reflection ends and
the associated Boltzmann weights. }
\end{center}


\noindent where $1$ and $2$ respectively denote the spin up and
down states. Each of the six bulk configurations is assigned a
statistical weight (or Boltzmann weight) $w_i$, while each of the
four reflection configurations is assigned a  weight $b_i$ (see Figs. 1 and 2).
Then the partition
function  of the model with a reflection end can be rewritten as
\begin{eqnarray}
Z=\sum\,{w_1}^{n_1}\,{w_2}^{n_2}\,{w_3}^{n_3}\,{w_4}^{n_4}\,{w_5}^{n_5}
 \,{w_6}^{n_6}\,b_1^{l_1}\,b_2^{l_2}\,b_3^{l_3}\,b_4^{l_4},\no
\end{eqnarray} where the summation is over all possible configurations with $n_i$
and $l_j$ being the number of the vertices of type $i$ and the number of the reflection end of
type $j$ respectively. The bulk
Boltzmann weights  which we consider here have $Z_2$-symmetry, i.e.,
\begin{eqnarray}
  a\equiv w_1=w_2,\quad b\equiv w_3=w_4, \quad c\equiv
  w_5=w_6,
\end{eqnarray} and the variables $a,b,c$ satisfy a function
relation, or equivalently, the local Boltzmann weights $\{w_i\}$
can be parameterized by the matrix elements of the six-vertex
R-matrix $R$ (\ref{r-matrix}) as in  figure 3. At the same time, the weights
$\{b_i\}$ corresponding to the reflection end can be parameterized by the matrix
elements of the K-matrix $K$ (\ref{K-matrix})  as in figure 3.


\begin{center}
\includegraphics[width=0.8\textwidth]{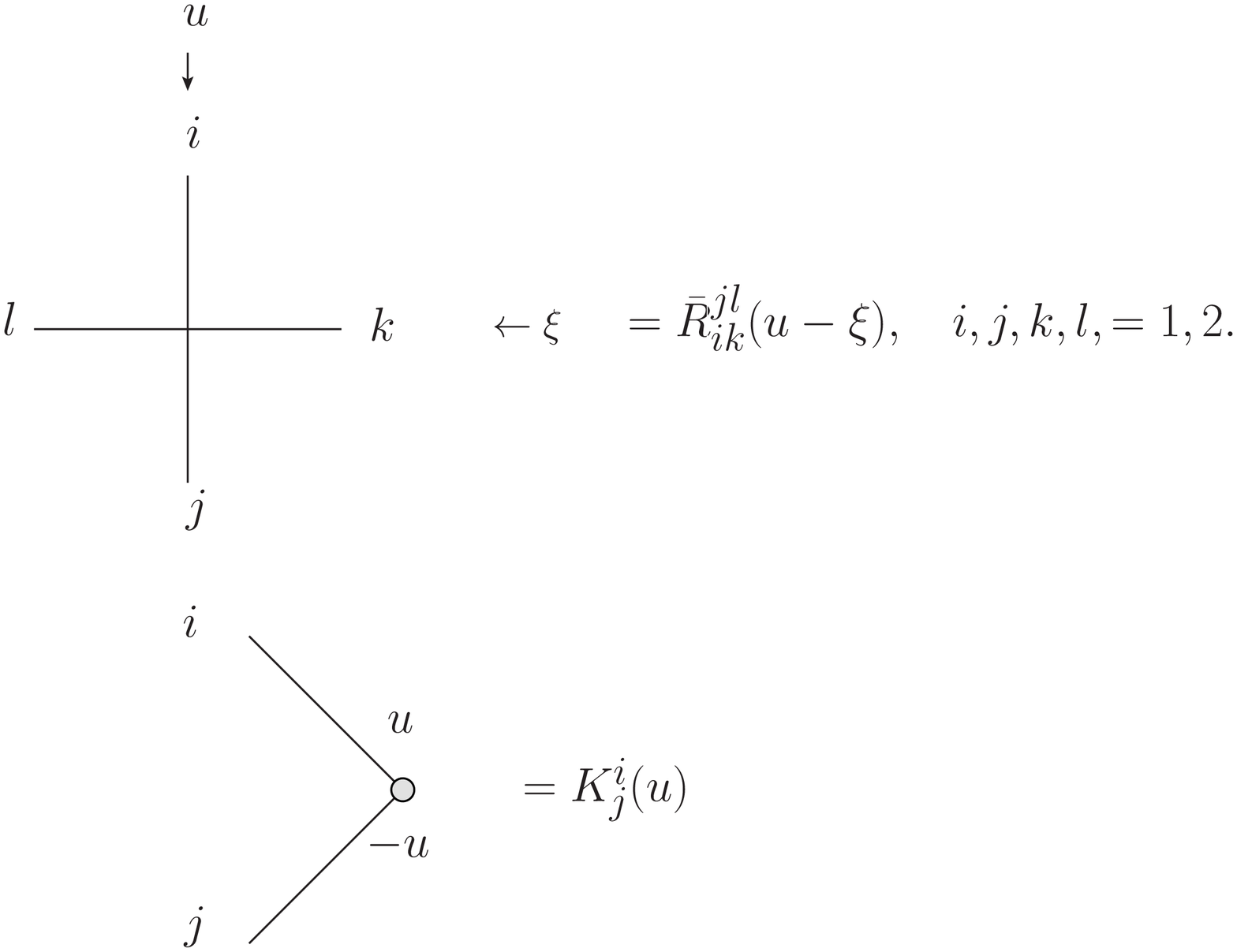}\\
{\small{\bf Figure 3.} The Boltzmann weights and elements of the six-vertex R-matrix and K-matrix. }
\end{center}


\noindent then the corresponding model is called the six-vertex
model with a reflection end. Therefore the
partition function of the model is give by
\begin{eqnarray}
Z=\sum\,{a}^{n_1+n_2}\,{b}^{n_3+n_4}\,{c}^{n_5+n_6}
 b_1^{l_1}\,b_2^{l_2}\,b_3^{l_3}\,b_4^{l_4}.\no
\end{eqnarray}

In order to parameterize the local bulk Boltzmann weights in terms of
the elements of the R-matrix, one needs to assign spectral
parameters $u$ and $\xi$ respectively to the vertical line and
horizontal line of each vertex of the lattice, as shown in  figure
3. In an inhomogeneous model, the statistical weights are
site-dependent. Hence two sets of spectral parameters
$\{u_{\alpha}\}$ and $\{\xi_i\}$ are needed, see figure 4. The
horizontal lines are enumerated by indices $1,\ldots,N$ with
spectral parameters $\{\xi_i\}$, while the vertical lines are
enumerated by indices $\bar{1},\ldots,\bar{N}$ with spectral
parameters $\{\bar{u}_{\alpha}\}$ (The $2N$ parameters $\{\bar{u}_{\alpha}\}$
are assigned as follow: $\bar{u}_{2i}=u_i$ and $\bar{u}_{2i+1}=-u_i$, as shown in figure 4.).
The DW boundary condition is
specified by four boundary states $|\Omega^{(2)}(\lambda)\rangle$,
$|\bar{\Omega}^{(1)}(\lambda)\rangle$,
$\langle\Omega^{(1)}(\lambda)|$ and
$\langle\bar{\Omega}^{(2)}(\lambda)|$ (the
definitions of the boundary states  will be given later,
see (\ref{Boundary-state-1})-(\ref{Boundary-state-4}) below).
These four states correspond to the particular choices of spin
states on the four boundaries of the lattice .

\vskip20mm
\begin{center}
\includegraphics[width=0.8\textwidth]{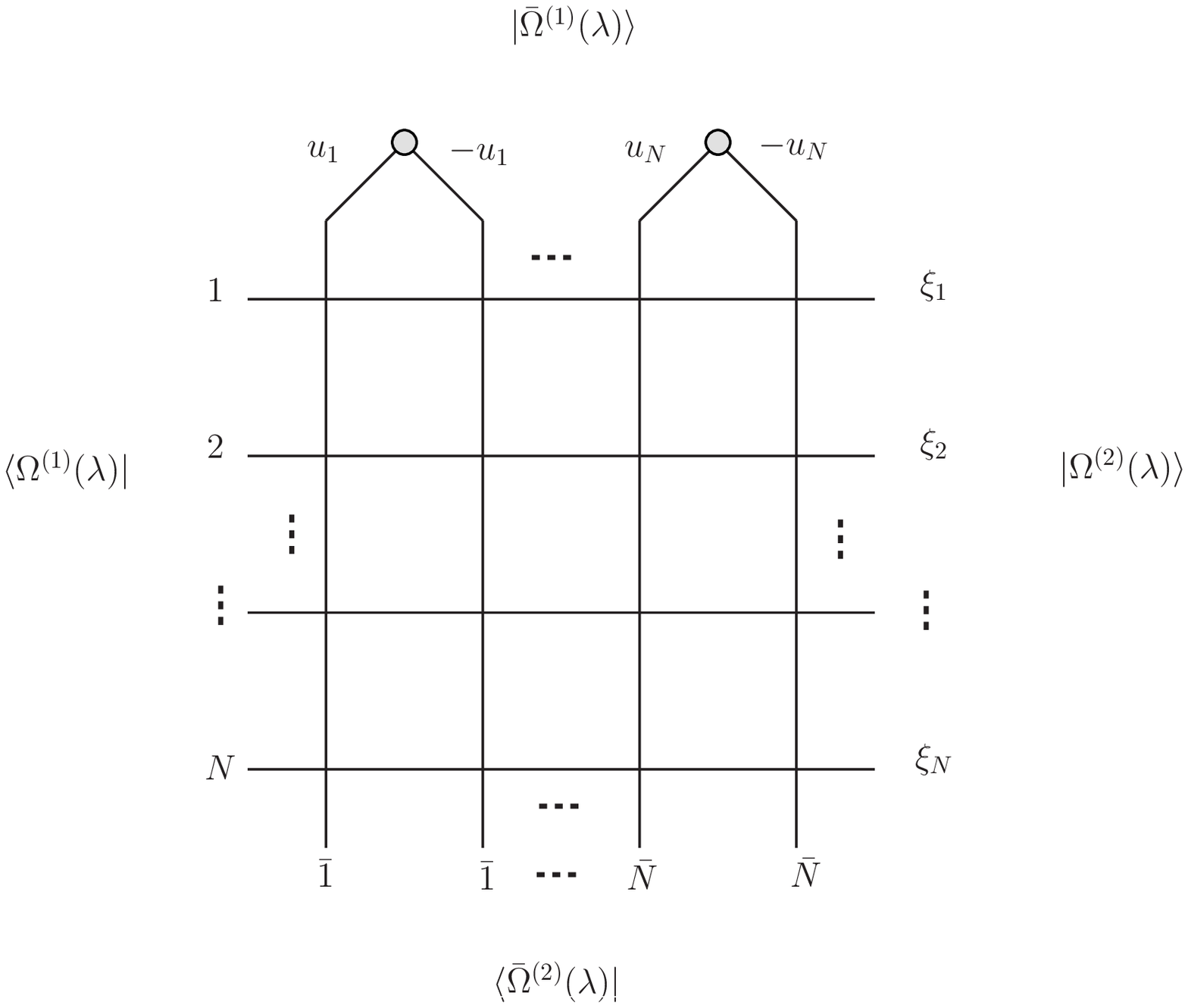}\\
{\small{\bf Figure 4.} The six-vertex model with a non-diagonal reflection end under the DW condition.}
\end{center}

Some remarks are in order. The boundary states not only depend
on the spectral parameters ($|\Omega^{(2)}(\lambda)\rangle$ and
$\langle\Omega^{(1)}(\lambda)|$ depend on
$\{\xi_i\}$, while $|\bar{\Omega}^{(1)}(\lambda)\rangle$ and
$\langle\bar{\Omega}^{(2)}(\lambda)|$ depend on
$\{u_{\alpha}\}$) but also on two continuous parameters
$\lambda_1$ and $\lambda_2$. However, after a diagonal similarity transformation generated by
${\rm Diag}(1,e^{-i(\l_1+\l_2)})$   and then
taking $\l_1\rightarrow +i\infty$, the corresponding boundary
states $|\Omega^{(2)}(\lambda)\rangle$ and
$\langle\bar{\Omega}^{(2)}( \lambda)|$ (or
$|\bar{\Omega}^{(1)}(\lambda)\rangle$ and
$\langle\Omega^{(1)}(\lambda)|$) become the state of
all spin down and its dual (or the state of all spin up and its
dual) up to some over-all scalar factors.  Moreover after the same similarity transformation and
taking the limit  the resulting K-matrix  becomes
the diagonal matrix solution of the reflection equation (\ref{RE-V}).  Hence the partition
function in the limit reduces to that of the six-vertex model
\cite{Tsu98}.

The partition function of the six-vertex model with a non-diagonal
reflection end specified by the generic K-matrix $K(u)$ (\ref{K-matrix})
under the DW boundary condition is a function of $2N+3$ variables $\{u_{\a}\}$,
$\{\xi_i\}$, $\l_1$, $\l_2$ and $\zeta$, which is denoted by
$Z_N(\{u_{\a}\};\{\xi_i\};\l;\zeta)$. Due to the fact that the local
Boltzmann weights of each vertex and reflection end of the lattice
are given by the matrix elements of the six-vertex R-matrix and the associated K-matrix (see figure 3),
the partition function can be expressed in terms of the product of the
R-matrices, the K-matrix and the four boundary states
\bea
 Z_N(\{u_{\a}\};\{\xi_i\};\l;\zeta)
    \hspace{-0.2truecm}&=&\hspace{-0.2truecm}\langle\Omega^{(1)}(\l)
    |\langle\bar{\Omega}^{(2)}(\l)|\no\\
 &&\times \R_{\bar{1},N}(u_1\hspace{-0.1truecm}-\hspace{-0.1truecm}\xi_N)\ldots
    \R_{\bar{1},1}(u_1\hspace{-0.1truecm}-\hspace{-0.1truecm}\xi_1)K_{\bar{1}}(u_1)
    \R_{1,\bar{1}}(u_1\hspace{-0.1truecm}+\hspace{-0.1truecm}\xi_1)\ldots
    \R_{N,\bar{1}}(u_1\hspace{-0.1truecm}+\hspace{-0.1truecm}\xi_N)\no\\
 &&\qquad\qquad\vdots\no\\
 &&\times \R_{\bar{N},N}(u_N\hspace{-0.1truecm}-\hspace{-0.1truecm}\xi_N)\ldots
    \R_{\bar{N},1}(u_N\hspace{-0.1truecm}-\hspace{-0.1truecm}\xi_1)K_{\bar{N}}(u_N)
    \R_{1,\bar{N}}(u_N\hspace{-0.1truecm}+\hspace{-0.1truecm}\xi_1)\ldots
    \R_{N,\bar{N}}(u_N\hspace{-0.1truecm}+\hspace{-0.1truecm}\xi_N)\no\\[8pt]
 && \times |\bar{\Omega}^{(1)}(\l)\rangle
    |\Omega^{(2)}(\l)\rangle.\label{PF}\eea
One can rearrange the product of the R-matrices in (\ref{PF}) in
terms of  a product  of the so-called double-row monodromy matrices
\bea
 \hspace{-1truecm}Z_N(\{u_{\a}\};\{\xi_i\};\l;\zeta)
 \hspace{-0.38truecm}&=&\hspace{-0.38truecm}\langle\Omega^{(1)}(\l)
    |\langle\bar{\Omega}^{(2)}(\l)|\,
    \mathbb{T}_{\bar{1}}(u_1)\ldots \mathbb{T}_{\bar{N}}(u_N)\,
    |\bar{\Omega}^{(1)}(\l)
    \rangle|\Omega^{(2)}(\l)\rangle,\label{PF-1}
\eea where the monodromy matrix $\mathbb{T}_{\bar{i}}(u)$ is given by
\bea
 \mathbb{T}_{\bar{i}}(u)\equiv \mathbb{T}_{\bar{i}}(u;\xi_1,\ldots,\xi_N;\zeta)&=&
 \R_{\bar{i},N}(u_i\hspace{-0.1truecm}-\hspace{-0.1truecm}\xi_N)\ldots
    \R_{\bar{i},1}(u_i\hspace{-0.1truecm}-\hspace{-0.1truecm}\xi_1)\no\\
   &&\quad\quad\times K_{\bar{i}}(u_i)\,
    \R_{1,\bar{i}}(u_i\hspace{-0.1truecm}+\hspace{-0.1truecm}\xi_1)\ldots
    \R_{N,\bar{i}}(u_i\hspace{-0.1truecm}+\hspace{-0.1truecm}\xi_N).
    \label{Monodromy-1}
\eea The double-row matrix $\mathbb{T}(u)$ has played an important role to construct the transfer matrix
for an open spin chain \cite{Skl88}. The QYBE (\ref{QYB}) of the R-matrix and the reflection equation (\ref{RE-V}) of the
K-matrix  give  rise to that the monodromy matrix $\mathbb{T}_{\bar{i}}(u)$ satisfy the following exchange relation
\bea
\R_{\bar{i},\bar{j}}(u_i-u_j)\,\mathbb{T}_{\bar{i}}(u_i)\,\R_{\bar{j},\bar{i}}(u_i+u_j)\,
    \mathbb{T}_{\bar{j}}(u_j)=\mathbb{T}_{\bar{j}}(u_j)\,\R_{\bar{i},\bar{j}}(u_i+u_j)\,
    \mathbb{T}_{\bar{i}}(u_i)\,\R_{\bar{j},\bar{i}}(u_i-u_j).\label{RLL-1}
\eea

\subsection{The boundary states}

From the orthonormal basis $\{\e_i\}$ of $V$, we define
\bea
\hat{\imath}=\e_i-\overline{\e},~~\overline{\e}=
\frac{1}{2}\sum_{k=1}^{2}\e_k, \quad i=1,2,\qquad {\rm then}\,
\sum_{i=1}^2\hat{\imath}=0. \label{fundmental-vector} \eea Let
$\h$ be the Cartan subalgebra of $A_{1}$ and $\h^{*}$ be its dual.
A finite dimensional diagonalizable  $\h$-module is a complex
finite dimensional vector space $W$ with a weight decomposition
$W=\oplus_{\mu\in \h^*}W[\mu]$, so that $\h$ acts on $W[\mu]$ by
$x\,v=\mu(x)\,v$, $(x\in \h,\,v\in\,W[\mu])$. For example, the
non-zero weight spaces of the fundamental representation
$V_{\L_1}=\Cb^2=V$ are
\bea
 W[\hat{\imath}]=\Cb \e_i,~i=1,2.\label{Weight}
\eea

For a generic $m\in V$, define \bea m_i=\langle m,\e_i\rangle,
~~m_{ij}=m_i-m_j=\langle m,\e_i-\e_j\rangle,~~i,j=1,2.
\label{Def1}\eea Let $R(u,m)\in {\rm End}(V\otimes V)$ be the
R-matrix of the six-vertex SOS model, which is trigonometric limit
of the eight-vertex SOS model \cite{Bax82} given by
\bea
R(u;m)\hspace{-0.1cm}=\hspace{-0.1cm}
\sum_{i=1}^{2}R(u;m)^{ii}_{ii}E_{ii}\hspace{-0.1cm}\otimes\hspace{-0.1cm}
E_{ii}\hspace{-0.1cm}+\hspace{-0.1cm}\sum_{i\ne
j}^2\lt\{R(u;m)^{ij}_{ij}E_{ii}\hspace{-0.1cm}\otimes\hspace{-0.1cm}
E_{jj}\hspace{-0.1cm}+\hspace{-0.1cm}
R(u;m)^{ji}_{ij}E_{ji}\hspace{-0.1cm}\otimes\hspace{-0.1cm}
E_{ij}\rt\}, \label{R-matrix} \eea where $E_{ij}$ is the matrix
with elements $(E_{ij})^l_k=\d_{jk}\d_{il}$. The coefficient
functions are \bea
 &&R(u;m)^{ii}_{ii}=1,~~
   R(u;m)^{ij}_{ij}=\frac{\sin u\sin(m_{ij}-\eta)}
   {\sin(u+\eta)\sin(m_{ij})},~~i\neq j,\label{Elements1}\\
 && R(u;m)^{ji}_{ij}=\frac{\sin\eta\sin(u+m_{ij})}
    {\sin(u+\eta)\sin(m_{ij})},~~i\neq j,\label{Elements2}
\eea  and $m_{ij}$ is defined in (\ref{Def1}). The R-matrix
satisfies the dynamical (modified) quantum Yang-Baxter equation
(or the star-triangle relation) \cite{Bax82}
\begin{eqnarray}
&&R_{1,2}(u_1-u_2;m-\eta h^{(3)})R_{1,3}(u_1-u_3;m)
R_{2,3}(u_2-u_3;m-\eta h^{(1)})\no\\
&&\qquad =R_{2,3}(u_2-u_3;m)R_{1,3}(u_1-u_3;m-\eta
h^{(2)})R_{1,2}(u_1-u_2;m).\label{MYBE}
\end{eqnarray}
Here we have adopted the convention
\bea R_{1,2}(u,m-\eta h^{(3)})\,v_1\otimes
v_2 \otimes v_3=\lt(R(u,m-\eta\mu)\otimes {\rm id }\rt)v_1\otimes
v_2 \otimes v_3,\quad {\rm if}\, v_3\in W[\mu]. \label{Action}
\eea Moreover, one may check that the R-matrix satisfies  weight
conservation condition, \bea
  \lt[h^{(1)}+h^{(2)},\,R_{1,2}(u;m)\rt]=0,\label{Conservation}
\eea unitary condition, \bea
 R_{1,2}(u;m)\,R_{2,1}(-u;m)={\rm id}\otimes {\rm
 id},\label{Unitary}
\eea and crossing relation \bea
 R(u;m)^{kl}_{ij}=\varepsilon_{l}\,\varepsilon_{j}
   \frac{\sin(u)\sin((m-\eta\hat{\imath})_{21})}
   {\sin(u+\eta)\sin(m_{21})}R(-u-\eta;m-\eta\hat{\imath})
   ^{\bar{j}\,k}_{\bar{l}\,i},\label{Crossing}
\eea where
\bea \varepsilon_{1}=1,\,\varepsilon_{2}=-1,\quad {\rm
and}\,\, \bar{1}=2,\,\bar{2}=1.\label{Parity} \eea

Define the following functions: $\theta^{(1)}(u)=e^{-iu}$, $
\theta^{(2)}(u)=1$. Let us introduce two intertwiners which are
$2$-component  column vectors $\phi_{m,m-\eta\hat{\jmath}}(u)$
labelled by $\hat{1},\,\hat{2}$. The $k$-th element of
$\phi_{m,m-\eta\hat{\jmath}}(u)$ is given by \bea
\phi^{(k)}_{m,m-\eta\hat{\jmath}}(u)=\theta^{(k)}(u+2m_j).\label{Intvect}\eea
Explicitly, \bea \phi_{m,m-\eta\hat{1}}(u)=
\lt(\begin{array}{c}e^{-i(u+2m_1)}\\1\end{array}\rt),\qquad
\phi_{m,m-\eta\hat{2}}(u)=
\lt(\begin{array}{c}e^{-i(u+2m_2)}\\1\end{array}\rt).\eea
Obviously, the two intertwiner vectors
$\phi_{m,m-\eta\hat{\imath}}(u)$ are linearly {\it independent}
for a generic $m\in V$.

 Using the intertwiner vectors, one can derive the following face-vertex
correspondence relation \cite{Cao03}\bea &&\R_{1,2}(u_1-u_2)
\phi^1_{m,m-\eta\hat{\imath}}(u_1)
\phi^2_{m-\eta\hat{\imath},m-\eta(\hat{\imath}+\hat{\jmath})}(u_2)
\no\\&&~~~~~~= \sum_{k,l}R(u_1-u_2;m)^{kl}_{ij}
\phi^1_{m-\eta\hat{l},m-\eta(\hat{l}+\hat{k})}(u_1)
\phi^2_{m,m-\eta\hat{l}}(u_2). \label{Face-vertex} \eea  Then the
QYBE (\ref{QYB}) of the vertex-type R-matrix $\R(u)$ is equivalent
to the dynamical Yang-Baxter equation (\ref{MYBE}) of the SOS
R-matrix $R(u,m)$. For a generic $m$, we can introduce other types
of intertwiners $\bar{\phi},~\tilde{\phi}$ which  are both row
vectors and satisfy the following conditions, \bea
  &&\bar{\phi}_{m,m-\eta\hat{\mu}}(u)
     \,\phi_{m,m-\eta\hat{\nu}}(u)=\d_{\mu\nu},\quad
     \tilde{\phi}_{m+\eta\hat{\mu},m}(u)
     \,\phi_{m+\eta\hat{\nu},m}(u)=\d_{\mu\nu},\label{Int2}\eea
{}from which one  can derive the relations,
\begin{eqnarray}
&&\sum_{\mu=1}^2\phi_{m,m-\eta\hat{\mu}}(u)\,
 \bar{\phi}_{m,m-\eta\hat{\mu}}(u)={\rm id},\label{Int3}\\
&&\sum_{\mu=1}^2\phi_{m+\eta\hat{\mu},m}(u)\,
 \tilde{\phi}_{m+\eta\hat{\mu},m}(u)={\rm id}.\label{Int4}
\end{eqnarray}

One may verify that the K-matrices $K(u)$ given by
(\ref{K-matrix}) can be expressed in terms
of the intertwiners and {\it diagonal\/} matrices $\K(\l|u)$
as follows
\bea &&K(u)^s_t=
\sum_{i,j}\phi^{(s)}_{\l-\eta(\hat{\imath}-\hat{\jmath}),
~\l-\eta\hat{\imath}}(u)
\K(\l|u)^j_i\bar{\phi}^{(t)}_{\l,~\l-\eta\hat{\imath}}(-u).
\label{K-F-1}
\eea Here the {\it
diagonal\/} matrix $\K(\l|u)$ is given
by \bea
&&\K(\l|u)\equiv{\rm Diag}(k(\l|u)_1,\,k(\l|u)_2)={\rm
Diag}(\frac{\sin(\l_1+\zeta-u)}{\sin(\l_1+\zeta+u)},\,
\frac{\sin(\l_2+\zeta-u)}{\sin(\l_2+\zeta+u)}).\label{K-F-3}
\eea
Although the vertex type K-matrix $K^{-}(u)$ given by
(\ref{K-matrix}) is  generally non-diagonal,
after the face-vertex transformation (\ref{K-F-1}), the face type
counterpart $\K(\l|u)$  becomes  diagonal. This
fact enabled the authors to apply the generalized algebraic Bethe
ansatz method developed in \cite{Yan04} for SOS type integrable
models to diagonalize the transfer matrix of the open XXZ chain with non-diagonal terms
\cite{Yan04-1,Yan07}.

Now we are in the position to construct the boundary states to specify the DW boundary
condition of the six-vertex model with a non-diagonal reflection end, see figure 4. For
any vector $m\in V$, we introduce four states which live in the two N-tensor spaces of $V$
(one is indexed by $1,\ldots,N$ and the other is indexed by $\bar{1},\ldots,\bar{N}$) or their
dual spaces as follows:
\bea
 |\Omega^{(2)}(m)\rangle&=&
   \phi^{1}_{m,m-\eta\hat{2}}(\xi_1)\,
   \phi^{2}_{m-\eta\hat{2},m-2\eta\hat{2}}(\xi_2)\ldots
   \phi^{N}_{m-\eta (N-1)\hat{2},m-\eta N\hat{2}}(\xi_N),\label{Boundary-state-1}\\[8pt]
|\bar{\Omega}^{(1)}(m)\rangle&=&\phi^{\bar{1}}_{m-(N-2)\eta\hat{1},m-(N-2)\eta\hat{1}-\eta\hat{1}}(-u_1)
   \,\phi^{\bar{1}}_{m-(N-4)\eta\hat{1},m-(N-4)\eta\hat{1}-\eta\hat{1}}(-u_2)\no\\
&&\times\ldots\phi^{\bar{N}}_{m+N\eta\hat{1},m+N\eta\hat{1}-\eta \hat{1}}(-u_N),
  \label{Boundary-state-2}\\[8pt]
\langle\Omega^{(1)}(m)|&=&
 \tilde{\phi}^{1}_{m,m-\eta\hat{1}}(\xi_1)\,
 \tilde{\phi}^{2}_{m-\eta\hat{1},m-2\eta\hat{1}}(\xi_2)\ldots
 \tilde{\phi}^{N}_{m-\eta(N-1)\hat{1},m-\eta N\hat{1}}(\xi_N),\label{Boundary-state-3}\\[8pt]
\langle\bar{\Omega}^{(2)}(m)|&=&
 \tilde{\phi}^{\bar{1}}_{m-N\eta\hat{1},m-N\eta\hat{1}-\eta\hat{2}}(u_1)\,
 \tilde{\phi}^{\bar{1}}_{m-(N-2)\eta\hat{1},m-(N-2)\eta\hat{1}-\eta\hat{2}}(u_2)\no\\
&&\times \ldots\tilde{\phi}^{\bar{N}}_{m+\eta(N-2)\hat{1},m-\eta (N-2)\hat{1}-\eta\hat{2}}(u_N).\label{Boundary-state-4}
\eea The boundary states which have been used to define the DW
boundary condition  can be obtained through the above
states by special choices of $m$ and $i$ (for example,  $m$ is
specified to $\l$ which is related to the parameters of the K-matrix $K(u)$).
Then the DW partition
function $Z_N(\{u_{\a}\};\{\xi_i\};\l;\zeta)$ of the six-vertex model with a non-diagonal reflection end given by (\ref{PF})
becomes
\bea
 &&Z_N(\{u_{\a}\};\{\xi_i\};\l;\zeta)=\no\\
 &&\quad\quad \tilde{\phi}^1_{\l,\l\hspace{-0.04truecm}-\hspace{-0.04truecm}\eta
  \hat{1}}(\xi_1)\ldots \tilde{\phi}^N_{\l\hspace{-0.04truecm}-\hspace{-0.04truecm}
  (N\hspace{-0.04truecm}-\hspace{-0.04truecm}1)\eta \hat{1},\l\hspace{-0.04truecm}-\hspace{-0.04truecm}N\eta\hat{1}}(\xi_N)
  \,\tilde{\phi}^{\bar{1}}_{\l\hspace{-0.04truecm}-\hspace{-0.04truecm}N\eta\hat{1},\l\hspace{-0.04truecm}-\hspace{-0.04truecm}N\eta
  \hat{1}\hspace{-0.04truecm}-\hspace{-0.04truecm}\eta\hat{2}}(u_1)\ldots \tilde{\phi}^{\bar{N}}
  _{\l\hspace{-0.04truecm}+\hspace{-0.04truecm}(N\hspace{-0.04truecm}-\hspace{-0.04truecm}2)\eta\hat{1},
  \l\hspace{-0.04truecm}+\hspace{-0.04truecm}(N\hspace{-0.04truecm}-\hspace{-0.04truecm}2)\eta\hat{1}
  \hspace{-0.04truecm}-\hspace{-0.04truecm}\eta\hat{2}}(u_N)\no\\
 &&\quad\qquad\times R_{\bar{1},N}(u_1-\xi_N)\ldots R_{\bar{1},1}(u_1-\xi_1)\,
    K_{\bar{1}}(u_1)\,R_{1,\bar{1}}(u_1+\xi_1)\ldots R_{N,\bar{1}}(u_1+\xi_N)\no\\
 && \qquad\qquad\qquad\qquad \vdots\no\\
 &&\quad\qquad\times R_{\bar{N},N}(u_N-\xi_N)\ldots R_{\bar{N},1}(u_N-\xi_1)\,
    K_{\bar{N}}(u_N)\,R_{1,\bar{N}}(u_N+\xi_1)\ldots R_{N,\bar{N}}(u_N+\xi_N)
    \no\\
 &&\quad\qquad\times \phi^{1}_{\l,\l\hspace{-0.04truecm}-\hspace{-0.04truecm}\eta\hat{2}}(\xi_1)\ldots
    \phi^{N}_{\l\hspace{-0.04truecm}-\hspace{-0.04truecm}(N\hspace{-0.04truecm}-\hspace{-0.04truecm}1)\eta\hat{2},
    \l\hspace{-0.04truecm}-\hspace{-0.04truecm}N\eta\hat{2}}(\xi_N)\,
    \phi^{\bar{1}}_{\l\hspace{-0.04truecm}-\hspace{-0.04truecm}(N\hspace{-0.04truecm}-\hspace{-0.04truecm}2)\eta\hat{1},
    \l\hspace{-0.04truecm}-\hspace{-0.04truecm}(N\hspace{-0.04truecm}-\hspace{-0.04truecm}1)\eta\hat{1}}
    (\hspace{-0.08truecm}-\hspace{-0.04truecm}u_1\hspace{-0.08truecm})
    \ldots\phi^{\bar{N}}_{\l\hspace{-0.04truecm}+\hspace{-0.04truecm}N\eta\hat{1},\l
    \hspace{-0.04truecm}+\hspace{-0.04truecm}(N\hspace{-0.04truecm}-\hspace{-0.04truecm}1)\eta\hat{1}}
    (\hspace{-0.08truecm}-\hspace{-0.04truecm}u_N\hspace{-0.08truecm}).\no\\
 && \label{PF-2}
\eea


\section{Partition function in terms of the face type monodromy matrix}
\label{Face} \setcounter{equation}{0}

Let us introduce the face type one-row monodromy matrix
\bea
 T_{F}(l|u)&\equiv &T^{F}_{0,1\ldots N}(l|u)\no\\
 &=&R_{0,N}(u-\xi_N;l-\eta\sum_{i=1}^{N-1}h^{(i)})\ldots
    R_{0,2}(u-\xi_2;l-\eta h^{(1)})R_{0,1}(u-\xi_1;l),\no\\
 &=&\lt(\begin{array}{ll}T_F(l|u)^1_1&T_F(l|u)^1_2\\T_F(l|u)^2_1&
   T_F(l|u)^2_2\end{array}\rt)
    \label{Monodromy-face-1}
\eea where $l$ is a generic vector in $V$. The monodromy matrix
satisfies the face type quadratic exchange relation
\cite{Fel96,Hou03}. Applying $T_F(l|u)^i_j$ to an arbitrary vector
$|i_1,\ldots,i_N\rangle$ in the N-tensor product space $V^{\otimes
N}$ given by \bea
   |i_1,\ldots,i_N\rangle=\e^1_{i_1}\ldots
   \e^N_{i_N},\label{Vector-V}
\eea we have \bea
 T_F(l|u)^i_j|i_1,\ldots,i_N\rangle&\equiv&
    T_F(m;l|u)^i_j|i_1,\ldots,i_N\rangle\no\\
 &=&\sum_{\a_{N-1}\ldots\a_1}\sum_{i'_N\ldots i'_1}
 R(u-\xi_N;l-\eta\sum_{k=1}^{N-1}\hat{\imath}'_k)
   ^{i\,\,\,\,\,\,\,\,\,\,\,\,\,\,i'_N}_{\a_{N-1}\,i_N}\ldots\no\\
 &&\quad\quad\times R(u-\xi_2;l-\eta\hat{\imath}'_1)^{\a_2\,i'_2}_{\a_1\,\,i_2}
 R(u-\xi_1;l)^{\a_1\,i'_1}_{j\,\,\,\,i_1}
   \,\,|i'_1,\ldots,i'_N\rangle,\label{Monodromy-face-2}
\eea where $m=l-\eta\sum_{k=1}^N\hat{\imath}_k$.

Now we compute the partition function $Z_N(\{u_{\a}\};\{\xi_i\};\l;\zeta)$ defined
by (\ref{PF}). The expression (\ref{PF-2}) implies that
\bea
 Z_N(\{u_{\a}\};\{\xi_i\};\l;\zeta)
     \hspace{-0.38truecm}&=&\hspace{-0.38truecm}\langle\Omega^{(1)}(\l)|\,
     \tilde{\phi}^{\bar{1}}_{\l\hspace{-0.04truecm}-\hspace{-0.04truecm}(N-1)\eta\hat{1}+\eta\hat{2},\l\hspace{-0.04truecm}-\hspace{-0.04truecm}(N-1)\eta
     \hat{1}}(u_1)\, \mathbb{T}_{\bar{1}}(u_1)\,
     \phi^{\bar{1}}_{\l\hspace{-0.04truecm}-\hspace{-0.04truecm}(N-2)\eta\hat{1},\l\hspace{-0.04truecm}-\hspace{-0.04truecm}(N-1)\eta
     \hat{1}}(-u_1) \no\\
 &&\qquad\qquad\vdots\no\\
 &&\times \tilde{\phi}^{\bar{N}}_{\l\hspace{-0.04truecm}+\hspace{-0.04truecm}(N-1)\eta\hat{1}+\eta\hat{2},\l\hspace{-0.04truecm}+\hspace{-0.04truecm}(N-1)\eta
     \hat{1}}(u_N)\, \mathbb{T}_{\bar{N}}(u_N)\,
     \phi^{\bar{N}}_{\l\hspace{-0.04truecm}+\hspace{-0.04truecm}N\eta\hat{1},\l\hspace{-0.04truecm}+\hspace{-0.04truecm}(N-1)\eta
     \hat{1}}(-u_N)\,|\Omega^{(2)}(\l)\rangle.\no\\
\eea
With the help of the crossing
relation (\ref{Crossing}), the face-vertex
correspondence relation (\ref{Face-vertex}) and the relations (\ref{Int2}),
following the
method developed in \cite{Yan04,Yan09}, we find that the partition function $Z_N(\{u_{\a}\};\{\xi_i\};\l;\zeta)$
can be expressed in terms of the face-type double-row monodromy operators as follows:
\bea
 &&Z_N(\{u_{\a}\};\{\xi_i\};\l;\zeta)=\langle 1,\ldots,1|\T^-_F(\l-2(M-1)\eta\hat{1},\l|u_1)^2_1
     \ldots\T^-_F(\l,\l|u_M)^2_1 \no\\
     &&\qquad\qquad\times \T^-_F(\l+2\eta\hat{1},\l|u_{M+1})^2_1 \ldots
     \T^-_F(\l+N\eta\hat{1},\l|u_N)^2_1 |2,\ldots,2\rangle.\label{Scalar-3}
\eea  The above double-row monodromy  matrix operator $\T^-_F(m,\l|u)^2_1$ is given
in terms of the one-row monodromy matrix operator $T_F(m;l|u)^i_j$
\cite{Yan09} as follow:
\bea
 &&\T^-_F(m,\l|u)^2_1=
 \frac{\sin(m_{21})}{\sin(\l_{21})}\prod_{k=1}^N
      \frac{\sin(u+\xi_k)}{\sin(u+\xi_k+\eta)}\no\\
 &&\,\quad\times\lt\{
      \frac{\sin(\l_1+\zeta-u)}{\sin(\l_1+\zeta+u)}
      T_F(m,\l|u)^2_1
      T_F(m+\eta\hat{2},\l+\eta\hat{2}|-u-\eta)^2_2\rt.\no\\
 &&\,\qquad-\lt.
      \frac{\sin(\l_2+\zeta-u)}{\sin(\l_2+\zeta+u)}
      T_F(m+2\eta\hat{2},\l|u)^2_2
      T_F(m+\eta\hat{1},\l+\eta\hat{1}|-u-\eta)^2_1\rt\}.\label{Expression-3}
 \eea

In the next section we  construct the Drinfeld twist (or
factorizing F-matrix) in the face picture for the six-vertex model with a
non-diagonal reflection end. In the resulting  F-basis, the
pseudo-particle creation operator $\T^{-}_F$ given by
(\ref{Expression-3})  takes  completely
symmetric and polarization free form. This polarization free form
allows us to construct the explicit expressions of the partition function
$Z_N(\{u_{\a}\};\{\xi_i\};\l;\zeta)$.


\section{ F-basis}
\label{F-basis} \setcounter{equation}{0}

In this section, we give the Drinfeld twist \cite{Dri83}
(factorizing F-matrix) on the $N$-fold tensor product space
$V^{\otimes N}$  and
the associated representations of the pseudo-particle creation/annihilation
operators in this basis.

\subsection{Factorizing Drinfeld twist $F$}
Let $ \mathcal{S}_N$ be the permutation group over indices
$1,\ldots,N$ and $\{\s_i|i=1,\ldots,N-1\}$ be the set of
elementary permutations in $\mathcal{S}_N$. For each elementary
permutation $\s_i$, we introduce the associated operator
$R^{\s_i}_{1\ldots N}$ on the quantum space
\bea
  R^{\s_i}_{1\ldots N}(l)\equiv R^{\s_i}(l)=R_{i,i+1}
    (\xi_i-\xi_{i+1}|l-\eta\sum_{k=1}^{i-1}h^{(k)}),\label{Fundamental-R-operator}
\eea where $l$ is a generic vector in $V$. For any $\s,\,\s'\in
\mathcal{S}_N$, operator $R^{\s\s'}_{1\ldots N}$ associated with
$\s\s'$ satisfies the following composition law \cite{Zha06}(and references therein):
\bea
  R_{1\ldots N}^{\s\s'}(l)=R^{\s'}_{\s(1\ldots
  N)}(l)\,R^{\s}_{1\ldots N}(l).\label{Rule}
\eea Let $\s$ be decomposed in a minimal way in terms of
elementary permutations,
\bea
  \s=\s_{\b_1}\ldots\s_{\b_p}, \label{decomposition}
\eea where $\b_i=1,\ldots, N-1$ and the positive integer $p$ is
the length of $\s$. The composition law (\ref{Rule}) enables one
to obtain  operator $R^{\s}_{1\ldots N}$ associated with each
$\s\in\mathcal{S}_N $. The dynamical quantum Yang-Baxter equation
(\ref{MYBE}), weight conservation condition (\ref{Conservation})
and unitary condition (\ref{Unitary}) guarantee the uniqueness of
$R^{\s}_{1\ldots N}$. Moreover, one may check that
$R^{\s}_{1\ldots N}$ satisfies the following exchange relation
with the face type one-row monodromy matrix
(\ref{Monodromy-face-1}) \bea
  R^{\s}_{1\ldots N}(l)T^F_{0,1\ldots N}(l|u)=T^F_{0,\s(1\ldots N)}(l|u)
    R^{\s}_{1\ldots N}(l-\eta h^{(0)}),\quad\quad \forall\s\in
    \mathcal{S}_N.\label{Exchang-Face-1}
\eea

Now, we construct the face-type Drinfeld twist $F_{1\ldots
N}(l)\equiv F_{1\ldots N}(l;\xi_1,\ldots,\xi_N)$ \footnote{In this
paper, we adopt the convention: $F_{\s(1\ldots N)}(l)\equiv
F_{\s(1\ldots N)}(l;\xi_{\s(1)},\ldots,\xi_{\s(N)})$.} on the $N$-fold
tensor product space $V^{\otimes N}$, which  satisfies the
following three properties:
\bea
 &&{\rm I.\,\,\,\,lower-triangularity;}\\
 &&{\rm II.\,\,\, non-degeneracy;}\\
 &&{\rm III.\,factorizing \, property}:\,\,
 R^{\s}_{1\ldots N}(l)\hspace{-0.08truecm}=\hspace{-0.08truecm}
    F^{-1}_{\s(1\ldots N)}(l)F_{1\ldots N}(l), \,\,
 \forall\s\in  \mathcal{S}_N.\label{Factorizing}
\eea Substituting (\ref{Factorizing}) into the exchange relation
(\ref{Exchang-Face-1}) yields the following relation
\bea
 F^{-1}_{\s(1\ldots N)}(l)F_{1\ldots N}(l)T^F_{0,1\ldots N}(l|u)=
   T^F_{0,\s(1\ldots N)}(l|u)F^{-1}_{\s(1\ldots N)}(l-\eta h^{(0)})
   F_{1\ldots N}(l-\eta h^{(0)}).
\eea Equivalently,
\bea
 F_{1\ldots N}(l)T^F_{0,1\ldots N}(l|u)F^{-1}_{1\ldots N}(l-\eta h^{(0)})
   =F_{\s(1\ldots N)}(l)T^F_{0,\s(1\ldots N)}(l|u)
   F^{-1}_{\s(1\ldots N)}(l-\eta h^{(0)}).\label{Invariant}
\eea Let us introduce the twisted monodromy matrix
$\tilde{T}^F_{0,1\ldots N}(l|u)$ by \bea
 \tilde{T}^F_{0,1\ldots N}(l|u)&=&
  F_{1\ldots N}(l)T^F_{0,1\ldots N}(l|u)F^{-1}_{1\ldots N}(l-\eta
  h^{(0)})\no\\
  &=&\lt(\begin{array}{ll}\tilde{T}_F(l|u)^1_1&\tilde{T}_F(l|u)^1_2
  \\\tilde{T}_F(l|u)^2_1&
   \tilde{T}_F(l|u)^2_2\end{array}\rt).\label{Twisted-Mon-F}
\eea Then (\ref{Invariant}) implies that the twisted monodromy
matrix is symmetric under $\mathcal{S}_N$, namely, \bea
 \tilde{T}^F_{0,1\ldots N}(l|u)=\tilde{T}^F_{0,\s(1\ldots
 N)}(l|u), \quad \forall \s\in \mathcal{S}_N.
\eea

Define the F-matrix:
\bea
  F_{1\ldots N}(l)=\sum_{\s\in
     \mathcal{S}_N}\sum^2_{\{\a_j\}=1}\hspace{-0.22truecm}{}^*\,\,\,\,
     \prod_{j=1}^NP^{\s(j)}_{\a_{\s(j)}}
     \,R^{\s}_{1\ldots N}(l),\label{F-matrix}
\eea where $P^i_{\a}$ is the embedding of the project operator
$P_{\a}$ in the $i^{{\rm th}}$ space with matrix elements
$(P_{\a})_{kl}=\d_{kl}\d_{k\a}$. The sum $\sum^*$ in
(\ref{F-matrix}) is over all non-decreasing sequences of the
labels $\a_{\s(i)}$:
\bea
  && \a_{\s(i+1)}\geq \a_{\s(i)}\quad {\rm if}\quad \s(i+1)>\s(i),\no\\
  && \a_{\s(i+1)}> \a_{\s(i)}\quad {\rm if}\quad
  \s(i+1)<\s(i).\label{Condition}
\eea From (\ref{Condition}), $F_{1\ldots N}(l)$ obviously is a
lower-triangular matrix. Moreover, the F-matrix is non-degenerate
because  all its diagonal elements are non-zero. It was shown \cite{Yan09}
that the F-matrix also satisfies the
factorizing property (\ref{Factorizing}). Hence, the F-matrix
$F_{1\ldots N}(l)$ given by (\ref{F-matrix}) is the desirable
Drinfeld twist.

\subsection{Completely symmetric  representations}
Direct calculation shows \cite{Yan09}  that the
twisted operators $\tilde{T}_F(l|u)^j_i$ defined by
(\ref{Twisted-Mon-F}) indeed simultaneously  have the
following polarization free forms. Here we present the results for the
relevant operators for our purpose
\bea
 &&\tilde{T}_F(l|u)^2_2=\frac{\sin(l_{21}-\eta)}{\sin\lt(l_{21}-\eta+
     \eta\langle H,\e_1\rangle\rt)}\otimes_{i}
     \lt(\begin{array}{ll}\frac{\sin(u-\xi_i)}{\sin(u-\xi_i+\eta)}&\\
     &1\end{array}\rt)_{(i)},\\[6pt]
 &&\tilde{T}_F(l|u)^2_1=\sum_{i=1}^N\frac{\sin\eta
     \sin(u\hspace{-0.08truecm}-\hspace{-0.08truecm}\xi_i
     \hspace{-0.08truecm}+\hspace{-0.08truecm}l_{12})}
     {\sin(u\hspace{-0.08truecm}-\hspace{-0.08truecm}\xi_i
     \hspace{-0.08truecm}+\hspace{-0.08truecm}\eta)\sin l_{12}} E_{12}^i\otimes_{j\neq i}
     \lt(\begin{array}{ll}\frac{\sin(u-\xi_j)\sin(\xi_i-\xi_j+\eta)}{\sin(u-\xi_j+\eta)\sin(\xi_i-\xi_j)}&\\
     &1\end{array}\rt)_{(j)}.
\eea Applying  the above operators
to the arbitrary  state $|i_1,\ldots,i_N\rangle$ given by
(\ref{Vector-V}) leads to
\bea
 &&\tilde{T}_F(m,l|u)^2_2=\frac{\sin(l_{21}-\eta)}{\sin\lt(l_{2}-m_1-\eta\rt)}
     \otimes_{i}
     \lt(\begin{array}{ll}\frac{\sin(u-\xi_i)}{\sin(u-\xi_i+\eta)}&\\
     &1\end{array}\rt)_{(i)},\\[6pt]
 &&\tilde{T}_F(m,l|u)^2_1=\sum_{i=1}^N
     \frac{\sin\eta
     \sin(u-\xi_i+l_{12})}{\sin(u-\xi_i+\eta)\sin l_{12}}\no\\
 &&\quad\quad\quad\quad\quad\quad
     \times    E_{12}^i \otimes_{j\neq i}
     \lt(\begin{array}{ll}\frac{\sin(u-\xi_j)\sin(\
     \xi_i-\xi_j+\eta)}{\sin(u-\xi_j+\eta)\sin(\xi_i-\xi_j)}&\\
     &1\end{array}\rt)_{(j)}.
\eea It then follows that the  pseudo-particle creation
operator (\ref{Expression-3}) in the
F-basis  has the following completely symmetric
polarization free form:
\bea
 &&\tilde{\T}^-_F(m,\l|u)^2_1=\frac{\sin m_{12}}{\sin(m_1-\l_2)}
   \prod_{k=1}^N\frac{\sin(u+\xi_k)}{\sin(u+\xi_k+\eta)}\no\\
  &&\quad\quad\times \sum_{i=1}^N\frac{\sin(\l_1+\zeta-\xi_i)\sin(\l_2+\zeta+\xi_i)\sin2u \sin\eta}
   {\sin(\l_1+\zeta+u)\sin(\l_2+\zeta+u)\sin(u-\xi_i+\eta)\sin(u+\xi_i)}\no\\[6pt]
  &&\quad\quad\quad\quad\quad\quad \times
   E_{12}^i\otimes_{j\neq i}\lt(\begin{array}{ll}
   \frac{\sin(u-\xi_j)\sin(u+\xi_j+\eta)\sin(\xi_i-\xi_j+\eta)}
   {\sin(u-\xi_j+\eta)\sin(u+\xi_j)\sin(\xi_i-\xi_j)}&\\
   &1\end{array}\rt)_{(j)}.\label{Creation-operator-1}
\eea


\section{Determinant representation of the partition function}
\label{F} \setcounter{equation}{0}

In this section we  compute  the DW partition function $Z_N(\{u_{\a}\};\{\xi_i\};\l;\zeta)$ from its expression (\ref{Scalar-3})
and the expansion  of the twisted operator
$\tilde{\T}^-_F(m,\l|u)^2_1$ (\ref{Creation-operator-1}) given in the previous section.

\subsection{Symmetric expression of the partition function}
From the definitions of the F-matrix
$F_{1\ldots N}(l)$ given by (\ref{F-matrix}), we can show that the state
$|2,\ldots,2\rangle$ and the dual state
$\langle 1,\ldots,1|$  are invariant under the action of $F_{1\ldots N}(l)$, namely,
\bea
 F_{1\ldots N}(l)|2,\ldots,2\rangle&=&|2,\ldots,2\rangle,\label{Invariant-1}\\
 \langle 1,\ldots,1|F_{1\ldots N}(l)&=&\langle 1,\ldots,1|.
 \label{Invariant-2}
\eea Hence the DW partition function $Z_N(\{u_{\a}\};\{\xi_j\};\l;\zeta)$
can be expressed in terms of the twisted operator $\tilde{\T}^-_F(m,\l|u)^2_1$ as follow
\bea
 Z_N(\{u_{\a}\};\{\xi_i\};\l;\zeta)&=&\hspace{-0.36truecm}\langle 1,\ldots,1|\T^-_F(\l\hspace{-0.1truecm}-\hspace{-0.1truecm}
     (N\hspace{-0.1truecm}-\hspace{-0.1truecm}2)\eta\hat{1},\l|u_1)^2_1
     \ldots\T^-_F(\l\hspace{-0.1truecm}+\hspace{-0.1truecm}N\eta\hat{1},\l|u_N)^2_1 |2,\ldots,2\rangle\no\\
 &=&\hspace{-0.36truecm}\langle 1,\ldots,1|\,F_{1\ldots N}(\l-N\eta\hat{1})\,\T^-_F(\l-(N-2)\eta\hat{1},\l|u_1)^2_1\ldots\no\\
 &&\times \T^-_F(\l+N\eta\hat{1},\l|u_N)^2_1\, F^{-1}_{1\ldots N}(\l+N\eta\hat{1})\,|2,\ldots,2\rangle\no\\
 &=&\hspace{-0.36truecm}\langle 1,\ldots,1|\tilde{\T}^-_F(\l\hspace{-0.1truecm}-\hspace{-0.1truecm}
     (N\hspace{-0.1truecm}-\hspace{-0.1truecm}2)\eta\hat{1},\l|u_1)^2_1
     \ldots\tilde{\T}^-_F(\l\hspace{-0.1truecm}+\hspace{-0.1truecm}N\eta\hat{1},\l|u_N)^2_1 |2,\ldots,2\rangle.\no
\eea Substituting the polarization free expression (\ref{Creation-operator-1}) of the twisted operator $\tilde{\T}^-_F(m,\l|u)^2_1$
into the above equation, we have {\small
\bea
&& \hspace{-0.46truecm}Z_N(\{u_{\a}\};\{\xi_i\};\l;\zeta)=
     \prod_{k=1}^M\frac{\sin(\l_{12}+2k\eta)\sin(\l_{12}-2k\eta+\eta)}
     {\sin(\l_{12}+k\eta)\sin(\l_{12}-k\eta+\eta)}
     \prod_{l=1}^N\prod_{i=1}^N\frac{\sin(u_i+\xi_l)}{\sin(u_i+\xi_l+\eta)}
     \langle 1,\ldots,1|\no\\[8pt]
  &&\hspace{-0.36truecm}\times\hspace{-0.1truecm}
     \sum_{i=1}^N\hspace{-0.1truecm}\frac{\sin(\l_1\hspace{-0.1truecm}+\hspace{-0.1truecm}\zeta
     \hspace{-0.1truecm}-\hspace{-0.1truecm}\xi_i)\sin(\l_2
     \hspace{-0.1truecm}+\hspace{-0.1truecm}\zeta\hspace{-0.1truecm}+\hspace{-0.1truecm}\xi_i)\sin2u_1 \sin\eta}
     {\sin(\hspace{-0.1truecm}\l_1\hspace{-0.1truecm}+\hspace{-0.1truecm}\zeta
     \hspace{-0.08truecm}+\hspace{-0.1truecm}u_1\hspace{-0.08truecm})\sin(\hspace{-0.08truecm}\l_2
     \hspace{-0.1truecm}+\hspace{-0.1truecm}\zeta\hspace{-0.1truecm}+\hspace{-0.1truecm}u_1\hspace{-0.08truecm})
     \sin(\hspace{-0.08truecm}u_1\hspace{-0.1truecm}-\hspace{-0.1truecm}\xi_i\hspace{-0.1truecm}+\hspace{-0.1truecm}\eta)
     \sin(\hspace{-0.08truecm}u_1\hspace{-0.1truecm}+\hspace{-0.1truecm}\xi_i\hspace{-0.08truecm})}
     E_{12}^i\hspace{-0.1truecm}\otimes_{j\neq i}\hspace{-0.1truecm}\lt(\hspace{-0.28truecm}\begin{array}{ll}
     \frac{\sin(u_1\hspace{-0.1truecm}-\xi_j)\sin(u_1\hspace{-0.1truecm}+\xi_j\hspace{-0.1truecm}+\eta)\sin(\xi_i\hspace{-0.1truecm}-\xi_j
     \hspace{-0.1truecm}+\eta)}
     {\sin(u_1\hspace{-0.1truecm}-\xi_j\hspace{-0.1truecm}+\eta)\sin(u_1\hspace{-0.1truecm}+\xi_j)\sin(\xi_i\hspace{-0.1truecm}-\xi_j)}&\\
     &1\end{array}\hspace{-0.28truecm}\rt)_{\hspace{-0.18truecm}(j)}\no\\
  &&\qquad\qquad\vdots\no\\
  &&\hspace{-0.36truecm}\times\hspace{-0.1truecm}
     \sum_{i=1}^N\hspace{-0.1truecm}\frac{\sin(\l_1\hspace{-0.1truecm}+\hspace{-0.1truecm}\zeta
     \hspace{-0.1truecm}-\hspace{-0.1truecm}\xi_i)\sin(\l_2
     \hspace{-0.1truecm}+\hspace{-0.1truecm}\zeta\hspace{-0.1truecm}+\hspace{-0.1truecm}\xi_i)\sin2u_N \sin\eta}
     {\sin(\hspace{-0.1truecm}\l_1\hspace{-0.1truecm}+\hspace{-0.1truecm}\zeta
     \hspace{-0.08truecm}+\hspace{-0.1truecm}u_N\hspace{-0.08truecm})\sin(\hspace{-0.08truecm}\l_2
     \hspace{-0.1truecm}+\hspace{-0.1truecm}\zeta\hspace{-0.1truecm}+\hspace{-0.1truecm}u_N\hspace{-0.08truecm})
     \sin(\hspace{-0.08truecm}u_N\hspace{-0.1truecm}-\hspace{-0.1truecm}\xi_i\hspace{-0.1truecm}+\hspace{-0.1truecm}\eta)
     \sin(\hspace{-0.08truecm}u_N\hspace{-0.1truecm}+\hspace{-0.1truecm}\xi_i\hspace{-0.08truecm})}
     E_{12}^i\hspace{-0.1truecm}\otimes_{j\neq i}\hspace{-0.1truecm}\lt(\hspace{-0.28truecm}\begin{array}{ll}
     \frac{\sin(u_N\hspace{-0.1truecm}-\xi_j)\sin(u_N\hspace{-0.1truecm}+\xi_j\hspace{-0.1truecm}+\eta)\sin(\xi_i\hspace{-0.1truecm}-\xi_j
     \hspace{-0.1truecm}+\eta)}
     {\sin(u_N\hspace{-0.1truecm}-\xi_j\hspace{-0.1truecm}+\eta)\sin(u_N\hspace{-0.1truecm}+\xi_j)\sin(\xi_i\hspace{-0.1truecm}-\xi_j)}&\\
     &1\end{array}\hspace{-0.28truecm}\rt)_{\hspace{-0.18truecm}(j)}\no\\[8pt]
  &&\hspace{-0.36truecm}\times |2,\ldots,2\rangle.\no
\eea}Expanding the last sum term of the above equation which corresponds to the contribution associated with
the spectral parameter $u_N$ yields {\small
\bea
&& \hspace{-0.46truecm}Z_N(\{u_{\a}\};\{\xi_i\};\l;\zeta)=
     \prod_{k=1}^M\frac{\sin(\l_{12}+2k\eta)\sin(\l_{12}-2k\eta+\eta)}
     {\sin(\l_{12}+k\eta)\sin(\l_{12}-k\eta+\eta)}
     \prod_{l=1}^N\prod_{i=1}^N\frac{\sin(u_i+\xi_l)}{\sin(u_i+\xi_l+\eta)}\no\\
  &&\hspace{-0.36truecm}\times\hspace{-0.1truecm}
     \sum_{i=1}^N\hspace{-0.1truecm}\frac{\sin(\l_1\hspace{-0.1truecm}+\hspace{-0.1truecm}\zeta
     \hspace{-0.1truecm}-\hspace{-0.1truecm}\xi_i)\sin(\l_2
     \hspace{-0.1truecm}+\hspace{-0.1truecm}\zeta\hspace{-0.1truecm}+\hspace{-0.1truecm}\xi_i)\sin2u_N \sin\eta}
     {\sin(\hspace{-0.1truecm}\l_1\hspace{-0.1truecm}+\hspace{-0.1truecm}\zeta
     \hspace{-0.08truecm}+\hspace{-0.1truecm}u_N\hspace{-0.08truecm})\sin(\hspace{-0.08truecm}\l_2
     \hspace{-0.1truecm}+\hspace{-0.1truecm}\zeta\hspace{-0.1truecm}+\hspace{-0.1truecm}u_N\hspace{-0.08truecm})
     \sin(\hspace{-0.08truecm}u_N\hspace{-0.1truecm}-\hspace{-0.1truecm}\xi_i\hspace{-0.1truecm}+\hspace{-0.1truecm}\eta)
     \sin(\hspace{-0.08truecm}u_N\hspace{-0.1truecm}+\hspace{-0.1truecm}\xi_i\hspace{-0.08truecm})}
     \prod_{l=1}^{N-1}\frac{\sin(u_l-\xi_i)\sin(u_l+\xi_i+\eta)}{\sin(u_l-\xi_i+\eta)\sin(u_l+\xi_i)}\no\\
  &&\hspace{-0.16truecm}\times\hspace{-0.1truecm}\prod_{j\neq i}\frac{\sin(\xi_j-\xi_i+\eta)}{\sin(\xi_j-\xi_i)}
      \,\langle 1,\ldots,1|\no\\[8pt]
  &&\hspace{-0.16truecm}\times\hspace{-0.1truecm}
     \sum_{l\neq i}^N\hspace{-0.1truecm}\frac{\sin(\l_1\hspace{-0.1truecm}+\hspace{-0.1truecm}\zeta
     \hspace{-0.1truecm}-\hspace{-0.1truecm}\xi_l)\sin(\l_2
     \hspace{-0.1truecm}+\hspace{-0.1truecm}\zeta\hspace{-0.1truecm}+\hspace{-0.1truecm}\xi_l)\sin2u_1 \sin\eta}
     {\sin(\hspace{-0.1truecm}\l_1\hspace{-0.1truecm}+\hspace{-0.1truecm}\zeta
     \hspace{-0.08truecm}+\hspace{-0.1truecm}u_1\hspace{-0.08truecm})\sin(\hspace{-0.08truecm}\l_2
     \hspace{-0.1truecm}+\hspace{-0.1truecm}\zeta\hspace{-0.1truecm}+\hspace{-0.1truecm}u_1\hspace{-0.08truecm})
     \sin(\hspace{-0.08truecm}u_1\hspace{-0.1truecm}-\hspace{-0.1truecm}\xi_l\hspace{-0.1truecm}+\hspace{-0.1truecm}\eta)
     \sin(\hspace{-0.08truecm}u_1\hspace{-0.1truecm}+\hspace{-0.1truecm}\xi_l\hspace{-0.08truecm})}
     E_{12}^i\hspace{-0.1truecm}\otimes_{j\neq l,i}\hspace{-0.1truecm}\lt(\hspace{-0.28truecm}\begin{array}{ll}
     \frac{\sin(u_1\hspace{-0.1truecm}-\xi_j)\sin(u_1\hspace{-0.1truecm}+\xi_j\hspace{-0.1truecm}+\eta)\sin(\xi_l\hspace{-0.1truecm}-\xi_j
     \hspace{-0.1truecm}+\eta)}
     {\sin(u_1\hspace{-0.1truecm}-\xi_j\hspace{-0.1truecm}+\eta)\sin(u_1\hspace{-0.1truecm}+\xi_j)\sin(\xi_l\hspace{-0.1truecm}-\xi_j)}&\\
     &1\end{array}\hspace{-0.28truecm}\rt)_{\hspace{-0.18truecm}(j)}\no\\
  &&\qquad\qquad\vdots\no\\
  &&\hspace{-0.36truecm}\times |2,\ldots,2\rangle.\no
\eea}Iterating the above procedure, we  obtain the
complete symmetric expression of the partition function
$Z_N(\{u_{\a}\};\{\xi_i\};\l;\zeta)$
\bea
Z_N(\{u_{\a}\};\{\xi_i\};\l;\zeta)&=&\prod_{k=1}^M\frac{\sin(\l_{12}+2k\eta)\sin(\l_{12}-2k\eta+\eta)}
  {\sin(\l_{12}+k\eta)\sin(\l_{12}-k\eta+\eta)}
  \prod_{l=1}^N\prod_{i=1}^N\frac{\sin(u_i+\xi_l)}{\sin(u_i+\xi_l+\eta)}\no\\
  &&\quad \times {\cal Z}_N(\{u_{\a}\};\{\xi_i\};\l;\zeta),
  \label{partition-1}
\eea where the normalized partition function ${\cal Z}_N(\{u_{\a}\};\{\xi_i\};\l;\zeta)$ is
\bea
   {\cal Z}_N(\{u_{\a}\};\{\xi_i\};\l;\zeta)\hspace{-0.38truecm}&=&\hspace{-0.38truecm}
   \sum_{\s\in\mathcal{S}_N}\prod_{n=1}^N\hspace{-0.08truecm}
   \lt\{\frac{\sin(\l_1\hspace{-0.08truecm}+\hspace{-0.08truecm}\zeta
   \hspace{-0.08truecm}-\hspace{-0.08truecm}\xi_{i_{\s(n)}})
   \sin(\l_2\hspace{-0.08truecm}+\hspace{-0.08truecm}\zeta
   \hspace{-0.08truecm}+\hspace{-0.08truecm}\xi_{i_{\s(n)}})
   \sin(2u_n)\sin\eta}
   {\sin(\l_1\hspace{-0.08truecm}+\hspace{-0.08truecm}\zeta
   \hspace{-0.08truecm}+\hspace{-0.08truecm}u_n)
   \sin(\l_2\hspace{-0.08truecm}+\hspace{-0.08truecm}\zeta
   \hspace{-0.08truecm}+\hspace{-0.08truecm}u_n)
   \sin(u_n\hspace{-0.08truecm}-\hspace{-0.08truecm}\xi_{i_{\s(n)}}
   \hspace{-0.08truecm}+\hspace{-0.08truecm}\eta)
   \sin(u_n\hspace{-0.08truecm}+\hspace{-0.08truecm}\xi_{i_{\s(n)}})}
   \rt.\no\\
   && \times \prod_{k>n}^N\lt.
   \frac{\sin(u_n-\xi_{i_{\s(k)}}) \sin(u_n+\xi_{i_{\s(k)}}+\eta) \sin(\xi_{i_{\s(n)}}-\xi_{i_{\s(k)}}+\eta)}
   {\sin(u_n-\xi_{i_{\s(k)}}+\eta) \sin(u_n+\xi_{i_{\s(k)}}) \sin(\xi_{i_{\s(n)}}-\xi_{i_{\s(k)}})}
   \rt\}.\label{Function-B-2}
\eea

\subsection{Recursive relation and the determinant representation}
From the expression (\ref{Function-B-2}) of the partition function ${\cal Z}_N(\{u_{\a}\};\{\xi_i\};\l;\zeta)$, it is easy to
check that the partition function is a symmetric function of $\{u_{\a}\}$ and $\{\xi_i\}$ separatively. Moreover,
we can derive that the partition function ${\cal Z}_N(\{u_{\a}\};\{\xi_i\};\l;\zeta)$ satisfy the following recursive
relation
\bea
 \hspace{-1.2truecm}{\cal Z}_N(\{u_{\a}\};\{\xi_i\};\l;\zeta)&=&\sum_{i=1}^N
     \frac{\sin(\l_1\hspace{-0.08truecm}+\hspace{-0.08truecm}\zeta
   \hspace{-0.08truecm}-\hspace{-0.08truecm}\xi_{i})
   \sin(\l_2\hspace{-0.08truecm}+\hspace{-0.08truecm}\zeta
   \hspace{-0.08truecm}+\hspace{-0.08truecm}\xi_{i})
   \sin(2u_N)\sin\eta}
   {\sin(\l_1\hspace{-0.08truecm}+\hspace{-0.08truecm}\zeta
   \hspace{-0.08truecm}+\hspace{-0.08truecm}u_N)
   \sin(\l_2\hspace{-0.08truecm}+\hspace{-0.08truecm}\zeta
   \hspace{-0.08truecm}+\hspace{-0.08truecm}u_N)
   \sin(u_N\hspace{-0.08truecm}-\hspace{-0.08truecm}\xi_{i}
   \hspace{-0.08truecm}+\hspace{-0.08truecm}\eta)
   \sin(u_N\hspace{-0.08truecm}+\hspace{-0.08truecm}\xi_{i})}\no\\
 &&\times \prod_{l=1}^{N-1}\frac{\sin(u_l-\xi_i)\sin(u_l+\xi_i+\eta)}
   {\sin(u_l-\xi_i+\eta)\sin(u_l+\xi_i)}\prod_{j\neq i}\frac{\sin(\xi_j-\xi_i+\eta)}
   {\sin(\xi_j-\xi_i)}\no\\
 &&\times {\cal Z}_{N-1}(\{u_{\a}\}_{\a\neq N};\{\xi_j\}_{j\neq i};\l;\zeta).\label{Recursive-relation}
\eea
One can show that the initial condition:
${\cal Z}_0(\{u_{\a}\};\{\xi_i\};\l;\zeta)=1$ and the  recursive relation (\ref{Recursive-relation})
{\it uniquely} determinate the partition function
${\cal Z}_N(\{u_{\a}\};\{\xi_i\};\l;\zeta)$ for any positive integer $N$. This fact allows us to
obtain the following determinant representation of the normalized partition function
${\cal Z}_N(\{u_{\a}\};\{\xi_i\};\l;\zeta)$:
\bea
{\cal Z}_N(\{u_{\a}\};\{\xi_i\};\l;\zeta)=
   \frac{\prod_{\a=1}^N\prod_{i=1}^N\sin(u_{\a}-\xi_i)\sin(u_{\a}+\xi_i+\eta)
   \,{\rm det}{\cal N}(\{u_{\a}\};\{\xi_i\})}
  {\prod_{\a>\b}\sin(u_{\a}\hspace{-0.1truecm}-\hspace{-0.1truecm}
  u_{\b})\sin(u_{\a}\hspace{-0.1truecm}+\hspace{-0.1truecm}u_{\b}
  \hspace{-0.1truecm}+\hspace{-0.1truecm}\eta)\prod_{k<l}
  \sin(\xi_k\hspace{-0.1truecm}-\hspace{-0.1truecm}\xi_l)\sin(\xi_k\hspace{-0.1truecm}+\hspace{-0.1truecm}\xi_l)},
  \label{partition-2}
\eea where the $N\times N$ matrix  ${\cal N}(\{u_{\a}\};\{\xi_i\})$ is given by
\bea
{\cal N}(\{u_{\a}\};\{\xi_i\})_{\a,j}&=&
  \frac{\sin\eta\sin(\l_1+\zeta-\xi_j)}
  {\sin(u_{\a}-\xi_j)\sin(u_{\a}+\xi_j+\eta)
  \sin(\l_1+\zeta+u_{\a})}\no\\
  &&\times \frac{\sin(\l_2+\zeta+\xi_j)\sin(2u_{\a})}
  {\sin(\l_2+\zeta+u_{\a})
  \sin(u_{\a}-\xi_j+\eta)
  \sin(u_{\a}+\xi_j)}.\label{Matrix}
\eea The proof of this representation is relegated to Appendix A.

Finally, we obtain the determinant representation of the partition function
$Z_N(\{u_{\a}\};\{\xi_i\};\l;\zeta)$ defined in (\ref{PF}) of the six-vertex model
with a non-diagonal reflection end under the DW boundary condition from the expression
(\ref{partition-1})
\bea
 Z_N(\{u_{\a}\};\{\xi_i\};\l;\zeta)&=&\prod_{k=1}^M\frac{\sin(\l_{12}+2k\eta)
     \sin(\l_{12}-2k\eta+\eta)}{\sin(\l_{12}+k\eta)\sin(\l_{12}-k\eta+\eta)}
     \prod_{l=1}^N\prod_{i=1}^N\frac{\sin(u_i+\xi_l)}{\sin(u_i+\xi_l+\eta)}\no\\
 &&\times \frac{\prod_{\a=1}^N\prod_{i=1}^N\sin(u_{\a}-\xi_i)\sin(u_{\a}+\xi_i+\eta)
   \,{\rm det}{\cal N}(\{u_{\a}\};\{\xi_i\})}
  {\prod_{\a>\b}\sin(u_{\a}\hspace{-0.1truecm}-\hspace{-0.1truecm}
  u_{\b})\sin(u_{\a}\hspace{-0.1truecm}+\hspace{-0.1truecm}u_{\b}
  \hspace{-0.1truecm}+\hspace{-0.1truecm}\eta)\prod_{k<l}
  \sin(\xi_k\hspace{-0.1truecm}-\hspace{-0.1truecm}\xi_l)\sin(\xi_k\hspace{-0.1truecm}+\hspace{-0.1truecm}\xi_l)},\no\\
  \label{partition-3}
\eea where the $N\times N$ matrix  ${\cal N}(\{u_{\a}\};\{\xi_i\})$ is given by (\ref{Matrix}).


\section{ Conclusions}
\label{C} \setcounter{equation}{0}

We have studied the partition function $Z_N(\{u_{\a}\};\{\xi_i\};\l;\zeta)$ of the six-vertex model with a non-diagonal reflection
end, where the corresponding K-matrix $K(u)$ given by (\ref{K-matrix}) is a generic non-diagonal solution of
the RE, under the DW boundary
condition. The  DW boundary condition is specified by four boundary states
(\ref{Boundary-state-1})-(\ref{Boundary-state-4}) which are two-parameter generalization of the
all-spin-down and all-spin-up states and their dual states. With the help of the F-basis provided by
the Drinfeld twist for the open XXZ spin chain with non-diagonal boundary
terms, we obtain the the complete symmetric expression (\ref{partition-1})-(\ref{Function-B-2}) of the
partition function. Such an explicit expression allows us to derive its
the recursive relation (\ref{Recursive-relation}). Solving the recursive relation,
we obtain the determinant representation (\ref{partition-3}) of the partition function $Z_N(\{u_{\a}\};\{\xi_i\};\l;\zeta)$.
The determinant representation of the partition function will play an important role to construct determinant representations of
scalar products between an
on-shell Bethe state and  a general state (or an off-shell Bethe state) of
the open XXZ chain with non-diagonal boundary
terms \cite{Yan10}.

\section*{Acknowledgements}
The financial supports from  the National Natural Science
Foundation of China (Grant Nos. 11075126 and 11031005), Australian Research Council
and the NWU Graduate Cross-discipline Fund (08YJC24)
are gratefully acknowledged.


\section*{Appendix A: Proof the determinant representation (\ref{partition-2}) }
\setcounter{equation}{0}
\renewcommand{\theequation}{A.\arabic{equation}}
In this appendix, we prove the determinant representation (\ref{partition-2}) of the normalized
partition function  ${\cal Z}_N(\{u_{\a}\};\{\xi_i\};\l;\zeta)$ defined in
(\ref{partition-1}). Let us introduce two series functions
$\{B_I(\{u_{\a}\};\{\xi_i\};\l;\zeta)\,|\,I=1,\ldots, N\}$ and
$\{F_I(\{u_{\a}\};\{\xi_i\};\l;\zeta)\,|\,I=1,\ldots,N\}$ which are given  respectively by
\bea
 \hspace{-1.2truecm} B_I(\{u_{\a}\};\{\xi_i\};\l;\zeta)&=&\hspace{-0.2truecm}
     \prod_{l=1}^I\frac{\sin(\l_1\hspace{-0.08truecm}+\hspace{-0.08truecm}\zeta
     \hspace{-0.08truecm}+\hspace{-0.08truecm}u_l)\sin(\l_2
     \hspace{-0.08truecm}+\hspace{-0.08truecm}\zeta\hspace{-0.08truecm}+\hspace{-0.08truecm}u_l)}
     {\sin(\l_1\hspace{-0.08truecm}+\hspace{-0.08truecm}\zeta\hspace{-0.08truecm}-\hspace{-0.08truecm}\xi_l)
     \sin(\l_2\hspace{-0.08truecm}+\hspace{-0.08truecm}\zeta\hspace{-0.08truecm}+\hspace{-0.08truecm}\xi_l)\sin2u_l}\,
     {\cal Z}_I(\{u_{\a}\};\{\xi_i\};\l;\zeta),
     \label{B-function}\\
  \hspace{-1.2truecm}F_I(\{u_{\a}\};\{\xi_i\};\l;\zeta)&=&\hspace{-0.28truecm}\frac{\prod_{\a=1}^I\prod_{j=1}^I\sin(u_{\a}-\xi_j)\sin(u_{\a}+\xi_j+\eta)}
     {\prod_{\a>\b}\sin(u_{\a}\hspace{-0.1truecm}-\hspace{-0.1truecm}
     u_{\b})\sin(u_{\a}\hspace{-0.1truecm}+\hspace{-0.1truecm}u_{\b}
     \hspace{-0.1truecm}+\hspace{-0.1truecm}\eta)\prod_{k<l}
     \sin(\xi_k\hspace{-0.1truecm}-\hspace{-0.1truecm}\xi_l)\sin(\xi_k\hspace{-0.1truecm}+\hspace{-0.1truecm}\xi_l)}
     \no\\[6pt]
  &&\hspace{-0.28truecm}\times {\rm det}\lt|\frac{\sin\eta}
     {\sin(u_{\a}\hspace{-0.1truecm}-\hspace{-0.1truecm}\xi_j)
     \sin(u_{\a}\hspace{-0.1truecm}+\hspace{-0.1truecm}\xi_j\hspace{-0.1truecm}+\hspace{-0.1truecm}\eta)
     \sin(u_{\a}\hspace{-0.1truecm}-\hspace{-0.1truecm}\xi_j\hspace{-0.1truecm}+\hspace{-0.1truecm}\eta)
     \sin(u_{\a}\hspace{-0.1truecm}+\hspace{-0.1truecm}\xi_j)}\rt|.\label{F-function}
\eea Then the proof of (\ref{partition-2}) is equivalent to the following identification
\bea
 B_I(\{u_{\a}\};\{\xi_i\};\l;\zeta)=F_I(\{u_{\a}\};\{\xi_i\};\l;\zeta),\quad {\rm for\,any\,positive\,integer\,\,I}.\label{Proof}
\eea We shall prove the above equation by induction.

\begin{itemize}

\item From direct calculation, we can show that (\ref{Proof}) holds for the case of $N=1$, namely,
\bea
 B_1(u_1;\xi_1;\l;\zeta)=F_1(u_1;\xi_1;\l;\zeta)=\frac{\sin\eta}{\sin(u_1-\xi_1+\eta)\sin(u_1+\xi_1)}.\no
\eea

\item Suppose that (\ref{Proof}) holds for the  case of $I\leq N-1$. We are to prove that it is satisfied also for
the case of $N$ as follows. It is easy to check that both $B_N(\{u_{\a}\};\{\xi_i\};\l;\zeta)$ and $F_N(\{u_{\a}\};\{\xi_i\};\l;\zeta)$
are symmetric functions of $\{u_{\a}\}$. Hence it is sufficient to prove that  as
function of $u_N$ they are equal to each other. The recursive relation (\ref{Recursive-relation})  of ${\cal Z}_N(\{u_{\a}\};\{\xi_i\};\l;\zeta)$
implies that $B_N(\{u_{\a}\};\{\xi_i\};\l;\zeta)$ satisfies the following
relation
\bea
 \hspace{-1.2truecm}B_N(\{u_{\a}\};\{\xi_i\};\l;\zeta)
    \hspace{-0.28truecm}&=&\hspace{-0.28truecm}\sum_{i=1}^N
    \hspace{-0.12truecm}\frac{\sin\eta}
   {\sin(u_N\hspace{-0.08truecm}-\hspace{-0.08truecm}\xi_{i}
   \hspace{-0.08truecm}+\hspace{-0.08truecm}\eta)
   \sin(u_N\hspace{-0.08truecm}+\hspace{-0.08truecm}\xi_{i})}\hspace{-0.12truecm}
   \prod_{l=1}^{N-1}\frac{\sin(u_l\hspace{-0.12truecm}-\hspace{-0.12truecm}\xi_i)
   \sin(u_l\hspace{-0.12truecm}+\hspace{-0.12truecm}\xi_i\hspace{-0.12truecm}+\hspace{-0.12truecm}\eta)}
   {\sin(u_l\hspace{-0.12truecm}-\hspace{-0.12truecm}\xi_i\hspace{-0.12truecm}+\hspace{-0.12truecm}\eta)
   \sin(u_l\hspace{-0.12truecm}+\hspace{-0.12truecm}\xi_i)}\no\\
 &&\times \prod_{j\neq i}
   \frac{\sin(\xi_j\hspace{-0.12truecm}-\hspace{-0.12truecm}\xi_i\hspace{-0.12truecm}+\hspace{-0.12truecm}\eta)}
   {\sin(\xi_j\hspace{-0.12truecm}-\hspace{-0.12truecm}\xi_i)}\,
   B_{N-1}(\{u_{\a}\}_{\a\neq N};\{\xi_j\}_{j\neq i};\l;\zeta).\label{Recursive-relation-B}
\eea The determinant representation of the function $F_N(\{u_{\a}\};\{\xi_i\};\l;\zeta)$ implies that it satisfies
the following recursive relation
\bea
 \hspace{-1.2truecm}F_N(\hspace{-0.08truecm}\{u_{\a}\};\hspace{-0.08truecm}\{\xi_i\};\hspace{-0.08truecm}\l;\zeta)
    \hspace{-0.28truecm}&=&\hspace{-0.28truecm}\sum_{i=1}^N
    \hspace{-0.12truecm}\frac{\sin\eta}
   {\sin(u_N\hspace{-0.08truecm}-\hspace{-0.08truecm}\xi_{i}
   \hspace{-0.08truecm}+\hspace{-0.08truecm}\eta)
   \sin(u_N\hspace{-0.08truecm}+\hspace{-0.08truecm}\xi_{i})}\hspace{-0.12truecm}
   \prod_{l=1}^{N-1}\frac{\sin(u_l\hspace{-0.12truecm}-\hspace{-0.12truecm}\xi_i)
   \sin(u_l\hspace{-0.12truecm}+\hspace{-0.12truecm}\xi_i\hspace{-0.12truecm}+\hspace{-0.12truecm}\eta)}
   {\sin(u_N\hspace{-0.12truecm}-\hspace{-0.12truecm}u_l)
   \sin(u_N\hspace{-0.12truecm}+\hspace{-0.12truecm}u_l+\eta)}\no\\
 &&\times \prod_{j\neq i}\hspace{-0.08truecm}
   \frac{\sin(u_N\hspace{-0.12truecm}-\hspace{-0.12truecm}\xi_j)
   \sin(u_N\hspace{-0.12truecm}+\hspace{-0.12truecm}\xi_j\hspace{-0.12truecm}+\hspace{-0.12truecm}\eta)}
   {\sin(\xi_j\hspace{-0.12truecm}-\hspace{-0.12truecm}\xi_i)\sin(\xi_j\hspace{-0.12truecm}+\hspace{-0.12truecm}\xi_i)}\,
   \hspace{-0.08truecm}F_{N-1}(\hspace{-0.08truecm}\{u_{\a}\}_{\a\neq N};
   \hspace{-0.08truecm}\{\xi_j\}_{j\neq i};\hspace{-0.08truecm}\l;\zeta).\label{Recursive-relation-F}
\eea The determinant representation (\ref{F-function}) and its recursive relation (\ref{Recursive-relation-F}) of the function
$F_N(\{u_{\a}\};\{\xi_i\};\l;\zeta)$ and the recursive relation  (\ref{Recursive-relation-B}) of $B_N(\{u_{\a}\};\{\xi_i\};\l;\zeta)$
imply that these two functions, as function of $u_N$, have the same simple poles located at \footnote{The determinant expression (\ref{F-function}) guarantees
that the apparent poles in (\ref{Recursive-relation-F}), which are located at $u_l,-u_l-\eta\,\,{\rm mod}(\sqrt{-1}\pi)$ for $l=1,\ldots,N-1$, do not
really be poles.}:
\bea
 \xi_i-\eta,\,-\xi_i\, \quad{\rm mod}(\sqrt{-1}\pi),\quad\quad i=1,\ldots,N.\label{Poles}
\eea Direct calculation shows that the residues of the two functions at each simple pole (\ref{Poles}) are indeed
the same. Moreover we
can show that
\bea
   B_N(\{u_{\a}\};\{\xi_i\};\l;\zeta)\lt|_{u_N\rightarrow\infty}\rt.=0=
   F_N(\{u_{\a}\};\{\xi_i\};\l;\zeta)\lt|_{u_N\rightarrow\infty}\rt.\no
\eea Thanks to the Liouville theorem, we can conclude that (\ref{Proof}) actually
holds for the case of $N$.
\end{itemize}
Finally we have completed the proof of (\ref{partition-2}).


\end{document}